\documentclass[useAMS,usenatbib]{mn2e}

\usepackage{natbib}
\usepackage{amsmath,amsfonts,amssymb}
\usepackage[dvips]{graphicx}
\usepackage{longtable}
\usepackage{float}
\usepackage{caption}
\usepackage{subfig}
\usepackage{txfonts}
\usepackage{multirow}
\usepackage[multiple]{footmisc}
\usepackage[dvips]{color}
\usepackage{comment}

\title[The 154~MHz radio sky observed by the Murchison Widefield Array]
  {The 154~MHz radio sky observed by the Murchison Widefield Array: noise, confusion and first source count
analyses}
\author[Franzen et al.]{T.~M.~O.~Franzen$^{1}$ \thanks{Email: thomas.franzen@curtin.edu.au},
C.~A.~Jackson$^{1,2}$,
A.~R.~Offringa$^{3}$,
R.~D.~Ekers$^{1,2}$,
R.~B.~Wayth$^{1,2}$,
\newauthor
G.~Bernardi$^{4,5,6}$,
J.~D.~Bowman$^{7}$,
F.~Briggs$^{8,2}$,
R.~J.~Cappallo$^{9}$,
A.~A.~Deshpande$^{10}$,
\newauthor
B.~M.~Gaensler$^{11,2,12}$,
L.~J.~Greenhill$^{6}$,
B.~J.~Hazelton$^{13}$,
M.~Johnston-Hollitt$^{14}$,
\newauthor
D.~L.~Kaplan$^{15}$,
C.~J.~Lonsdale$^{9}$,
S.~R.~McWhirter$^{9}$,
D.~A.~Mitchell$^{16,2}$,
M.~F.~Morales$^{13}$,
\newauthor
E.~Morgan$^{17}$,
J.~Morgan$^{1}$,
D.~Oberoi$^{18}$,
S.~M.~Ord$^{1,2}$,
T.~Prabu$^{10}$,
N.~Seymour$^{1}$,
\newauthor
N.~Udaya~Shankar$^{10}$,
K.~S.~Srivani$^{10}$,
R.~Subrahmanyan$^{10,2}$,
S.~J.~Tingay$^{1,2}$,
C.~M.~Trott$^{1,2}$,
\newauthor
R.~L.~Webster$^{19,2}$,
A.~Williams$^{1}$
and C.~L.~Williams$^{17}$
\\
$^{1}$International Centre for Radio Astronomy Research, Curtin University, Bentley, WA 6102, Australia\\
$^{2}$ARC Centre of Excellence for All-sky Astrophysics (CAASTRO)\\
$^{3}$Netherlands Institute for Radio Astronomy (ASTRON), PO Box 2, 7990 AA Dwingeloo, The Netherlands\\
$^{4}$SKA SA, 3rd Floor, The Park, Park Road, Pinelands, 7405, South Africa\\
$^{5}$Department of Physics and Electronics, Rhodes University, PO Box 94, Grahamstown 6140, South Africa\\
$^{6}$Harvard-Smithsonian Center for Astrophysics, 60 Garden Street, Cambridge, MA 02138, USA\\
$^{7}$School of Earth and Space Exploration, Arizona State University, Tempe, AZ 85287, USA\\
$^{8}$Research School of Astronomy and Astrophysics, Australian National University, Canberra, ACT 2611, Australia\\
$^{9}$MIT Haystack Observatory, Westford, MA 01886, USA\\
$^{10}$Raman Research Institute, Bangalore 560080, India\\
$^{11}$Sydney Institute for Astronomy, School of Physics, The University of Sydney, NSW 2006, Australia\\
$^{12}$Dunlap Institute for Astronomy and Astrophysics, University of Toronto, ON, M5S 3H4, Canada\\
$^{13}$Department of Physics, University of Washington, Seattle, WA 98195, USA\\
$^{14}$School of Chemical \& Physical Sciences, Victoria University of Wellington, Wellington 6140, New Zealand\\
$^{15}$Department of Physics, University of Wisconsin--Milwaukee, Milwaukee, WI 53201, USA\\
$^{16}$CSIRO Astronomy and Space Science (CASS), PO Box 76, Epping, NSW 1710, Australia\\
$^{17}$Kavli Institute for Astrophysics and Space Research, Massachusetts Institute of Technology, Cambridge, MA 02139, USA\\
$^{18}$National Centre for Radio Astrophysics, Tata Institute for Fundamental Research, Pune 411007, India\\
$^{19}$School of Physics, The University of Melbourne, Parkville, VIC 3010, Australia}

\date{Accepted ????. Received ????}
\pagerange{\pageref{firstpage}--\pageref{lastpage}}
\pubyear{2012}

\begin{document}
\maketitle
\label{firstpage}

\begin{abstract}

\noindent

We analyse a 154~MHz image made from a 12~h observation with the Murchison Widefield Array (MWA) to determine the noise contribution and behaviour of the source counts down to 30~mJy. The MWA image has a bandwidth of 30.72~MHz, a field-of-view within the
half-power contour of the primary beam of $570~\mathrm{deg}^{2}$, a resolution of 2.3~arcmin and contains 13,458 sources above $5 \sigma$. 
The rms noise in the centre of the image is $4-5~\mathrm{mJy/beam}$. The MWA counts are in excellent agreement with counts
from other instruments and are the most precise ever derived in the flux density range 30--200~mJy due to
the sky area covered. Using the deepest available source count data, we find that the MWA image is affected by sidelobe confusion noise at the $\approx 3.5$~mJy/beam level, due to incompletely-peeled and out-of-image sources, and classical confusion becomes apparent 
at $\approx 1.7$~mJy/beam. This work highlights that (i) further improvements in ionospheric calibration and deconvolution 
imaging techniques would be required to probe to the classical confusion limit and (ii) the shape of low-frequency source counts,
including any flattening towards lower flux densities, must be determined from deeper $\approx 150$~MHz surveys as it cannot be directly inferred from higher frequency data.

\end{abstract}


\section{Introduction}\label{Introduction}

Radio source counts embody information about the extragalactic source populations and their evolution (i.e. space density) over cosmic time as determined by \cite{longair1966} and many others since. Whilst bright sources are relatively easy to identify, they are rare; the vast bulk of radio continuum emission emanates from moderate and low-power extragalactic radio sources whose radio emission is due to a central Active Galactic Nucleus (AGN) and/or ongoing star formation. These sources are distributed over a large range of redshifts, and thus contribute to the source counts to low flux densities. Surveys at a wide range of radio frequencies probe the faint sky, although at low frequencies ($< 200$~MHz), large-area ($> 100~\mathrm{deg}^{2}$) surveys remain limited to confusion effects at the mJy level, mainly due to large instrumental beam sizes. The situation is expected to improve with the extensive baselines and sensitivity of the Low Frequency Array \citep[LOFAR;][]{van_haarlem2013} and Square Kilometre Array Low \citep{dewdney2012}, which should push this limit substantially fainter. 

Radio source counts can be used to derive the integrated sky brightness and the power spectrum signature of the 
extragalactic sources \citep[e.g.][]{condon2012,toffolatti1998}.
The typical sensitivity limit to which sources can be reliably extracted from a uniform survey is 5$\sigma$, where $\sigma$ is due to the combination of confusion and system noise. However, even in fairly low resolution images where the noise is dominated by classical confusion, survey data can be statistically probed below the 5$\sigma$ 
limit using the $P(D)$ \citep{scheuer1957} distribution analysis to deduce the probable underlying source count behaviour 
\citep[see e.g.][]{mitchell1985,condon2012}. The large field-of-view (FoV) of the Murchison Widefield Array (MWA) and the huge number of detected sources 
gives rise to potential sidelobe confusion in images.  Although we know that the deepest MWA images to date are confusion 
limited, the {\it relative} contribution of classical and sidelobe confusion is poorly determined: this makes it hard to statistically interpret survey data 
below the source detection threshold and to assess whether enhancements in the data processing, such as improved deconvolution techniques, have the 
potential to further reduce the noise.

Whilst our science driver is to determine the MWA radio source counts to probe the contributing extragalactic source populations and their evolution, these data are also important for analyses where these sources are considered contaminating foregrounds. A number of new instruments, including the MWA are seeking to detect the first global signals from the Epoch of Reionisation (EoR); these rely on direct foreground source subtraction or isolation of the composite foreground signal to isolate the much fainter EoR signal in the power spectra.

MWA EoR observations concentrate on fields selected at high Galactic latitude free of diffuse Galactic emission. There are two options to extracting the EoR signal from the foreground signals: (i) via direct foreground subtraction and (ii) via their statistical suppression within the power spectrum \citep[see e.g.][]{morales2004,harker2010,chapman2012,trott2012,parsons2014}. Both methods benefit from high-validity source catalogues, and for (ii), a significant extrapolation of the known source counts to model the behaviour of foreground sources to deep flux density limits to permit maximal analysis of the power spectrum.

In the absence of the availability of low frequency source counts, their behaviour has been deduced by extrapolating the counts at 1400~MHz, which are well determined to $\approx$10 $\mu$Jy. This approach was used by \cite{thyagarajan2013a} to estimate the level of foreground contamination expected in MWA EoR power spectra. However, adopting simple spectral index conversions is unreliable because the shape of the radio source counts changes with frequency due to the changing nature of the sources contributing to the counts at 1400 and 154~MHz, and the relative dominance of (any) flat-spectrum, beamed component(s) (see e.g. \cite{wall1994}, \cite{jackson1999} and references therein).  

In this paper, we use an image of one MWA EoR field (EoR0) to determine the 154~MHz source counts down to $\approx 30$~mJy.
We can probe the behaviour of classical and sidelobe confusion noise at $S < 30$~mJy by
comparing with other source count data: this approach allows us to determine that the classical confusion noise becomes apparent at $\approx 1.7$~mJy/beam
and the sidelobe confusion noise can be expressed as a Gaussian distribution with $\mathrm{rms} \approx 3.5$~mJy/beam. Given that the sidelobe confusion noise 
is larger than the classical confusion limit, we do not attempt to extrapolate the behaviour of the 154~MHz source counts. Instead we investigate how 
sensitive our estimates are to a flattening in the source count slope below 6~mJy. In conclusion, we discuss likely origins of sidelobe confusion in
MWA images and areas of future work.

\section{MWA instrument and noise characteristics}\label{MWA data}

The MWA is an interferometer comprised of 128 16-crossed-pair-dipole antenna `tiles' distributed over an area $\approx 3$~km in diameter. 
It operates at frequencies between 72 and 300~MHz, with an instantaneous bandwidth of 30.72~MHz.
It is located at the Murchison Radio-astronomy Observatory in Western Australia and is 
the low-frequency precursor telescope for the SKA.
We refer the reader to \cite{lonsdale2009} and \cite{tingay2013} for details of the technical design and specifications of the MWA. 
The primary science objectives of the MWA are detailed in \cite{bowman2013}.
Using a uniform image weighting scheme, the angular resolution at 154~MHz is approximately
2.5 by 2.2~$\mathrm{sec}(\delta + 26.7^{\circ})$~arcsec. Given the effective width ($\approx 4$~m) of 
the MWA's antenna tiles, the primary beam FWHM is $27^{\circ}$ at 154~MHz.
The excellent snapshot $uv$ coverage of the MWA, owing to the very large number (8128) of baselines,
and its huge FoV allow it to rapidly image large areas of sky.

A key science goal for the MWA is to search for the redshifted 21-cm emission from the
EoR in the early Universe. Several fields are being targeted with deep 
(accumulating up to 1000~h), pointed observations \citep[see e.g.][]{beardsley2013}. The 
confusion noise in these EoR images is worse 
than the thermal noise as we show in Section~\ref{Quantifying the classical and sidelobe confusion noise}, 
making them ideal for measuring confusion. They also cover a sufficiently large area of 
sky to allow the source counts to be measured accurately over a wide range of flux densities.

There are three contributions to the total noise in all MWA images: system noise, classical confusion and
sidelobe confusion, where we take sidelobe confusion to include calibration errors and smearing effects. In the remainder of this section, 
we briefly describe these in context of our current understanding of MWA observations.

\subsection{Thermal noise}

The Gaussian random noise term, $T_{\mathrm{sys}}$, is equal to $T_{\mathrm{sky}} + T_{\mathrm{rec}}$, 
where $T_{\mathrm{sky}}$ is the sky noise and $T_{\mathrm{rec}}$ the receiver noise. Given the low 
observing frequency of the MWA, $T_{\mathrm{sys}}$ is dominated by $T_{\mathrm{sky}}$, with a far lower 
contribution from $T_{\mathrm{rec}}$. The thermal noise contribution in real MWA data can be estimated
using an imaging mode with no confusion. Stokes $V$ data are ideal providing identical aperture plane
coverage and noise characteristics. In a single MWA 2~min snapshot at high Galactic latitude, for a 
central frequency of 154~MHz and a bandwidth of 30.72~MHz, the measured rms noise in uniformly weighted Stokes 
$V$ images is $\approx 16$~mJy/beam.

From \cite{tingay2013}, in theory, the naturally weighted sensitivity 
for the same integration time, central frequency and bandwidth is $\approx 5$~mJy/beam (this assumes $T_{\mathrm{sky}} = 350$~K
and $T_{\mathrm{rec}} = 50$~K). After accounting for a 2.1-fold loss in sensitivity
due to uniform weighting \citep{wayth2015} and a reduction of $\approx 25$ per cent in the bandwidth due
to flagged edge channels, the theoretical prediction is $\frac{2.1}{\sqrt{0.75}} \times 5~\mathrm{mJy/beam} \approx 12~\mathrm{mJy/beam}$,
which compares well with our measurement above.

\subsection{Classical confusion}

Classical confusion occurs when the surface density of faint extragalactic sources is high enough
to prevent the sources from being resolved by the array. The fluctuations in the image are 
due to the sum of all sources in the main lobe of the synthesised beam. Classical confusion only 
depends on the source counts and the synthesised beam area \citep{condon1974}. 

\cite{bernardi2009} used a power spectrum analysis to estimate the classical confusion noise in three 
$6 \times 6 ~\mathrm{deg}^{2}$ sky areas observed with the Westerbork Synthesis Radio Telescope at 150~MHz. 
They estimated the rms classical confusion noise at 150~MHz, $\sigma_{\mathrm{c}}$, to be 3~mJy/beam for a 2~arcmin beam.
Other analyses to estimate the MWA classical confusion limit 
have extrapolated higher frequency source counts given the paucity of deep 150~MHz source count data
and have adopted slightly different beam size estimates. Using the method described in \cite{thyagarajan2013a}, \cite{thyagarajan2013b} estimated
$\sigma_{\mathrm{c}}$ from extrapolation of the 1400~MHz counts by \cite{hopkins2003} to 150~MHz. Assuming 
$\alpha^{1400}_{150} = -0.78$ ($S \propto \nu^{\alpha}$) and a source subtraction limit of 5$\sigma$, they obtained $\sigma_{\mathrm{c}} = 3~\mathrm{mJy/beam}$ for a 2~arcmin beam. \cite{wayth2015} estimated $\sigma_{\mathrm{c}}$ from extrapolation of the 327~MHz counts
measured by \cite{wieringa1991} down to 4~mJy. Following \citeauthor{condon1974} and using a signal-to-noise threshold of 6,
\citeauthor{wayth2015} obtained $\sigma_{\mathrm{c}} = 2$~mJy/beam for a 2.4~arcmin beam. 

LOFAR EoR observations are probing the 115--190~MHz sky to $\approx 30~\mu \mathrm{Jy/beam}$ rms, 
although no deep extragalactic source catalogues are yet available. Observations with the Giant Metrewave Radio Telescope (GMRT)
by \cite{intema2011}, \cite{ghosh2012} and \cite{williams2013} probe the 153~MHz counts down to 6, 12 and 15~mJy, 
respectively. In Section~\ref{Quantifying the classical and sidelobe confusion noise}, we use these deep source counts to quantify the
classical confusion noise in the MWA data.

\subsection{Sidelobe confusion}

Sidelobe confusion introduces additional noise into an image due to imperfect source 
deconvolution within the image; i.e. by all sources below the source subtraction cut-off limit and also
from the array response to sources outside the imaged FoV.
The MWA array has an irregular layout
(i.e. station baselines are unique) and performs a huge number (8128) of correlations 
such that sidelobes from any single short observation will be randomly distributed and hard to distinguish 
from real sources or other noise elements.

\begin{figure}
 \begin{center}
  \includegraphics[scale=0.36, trim=2cm 6cm 0cm 7cm, angle=270]{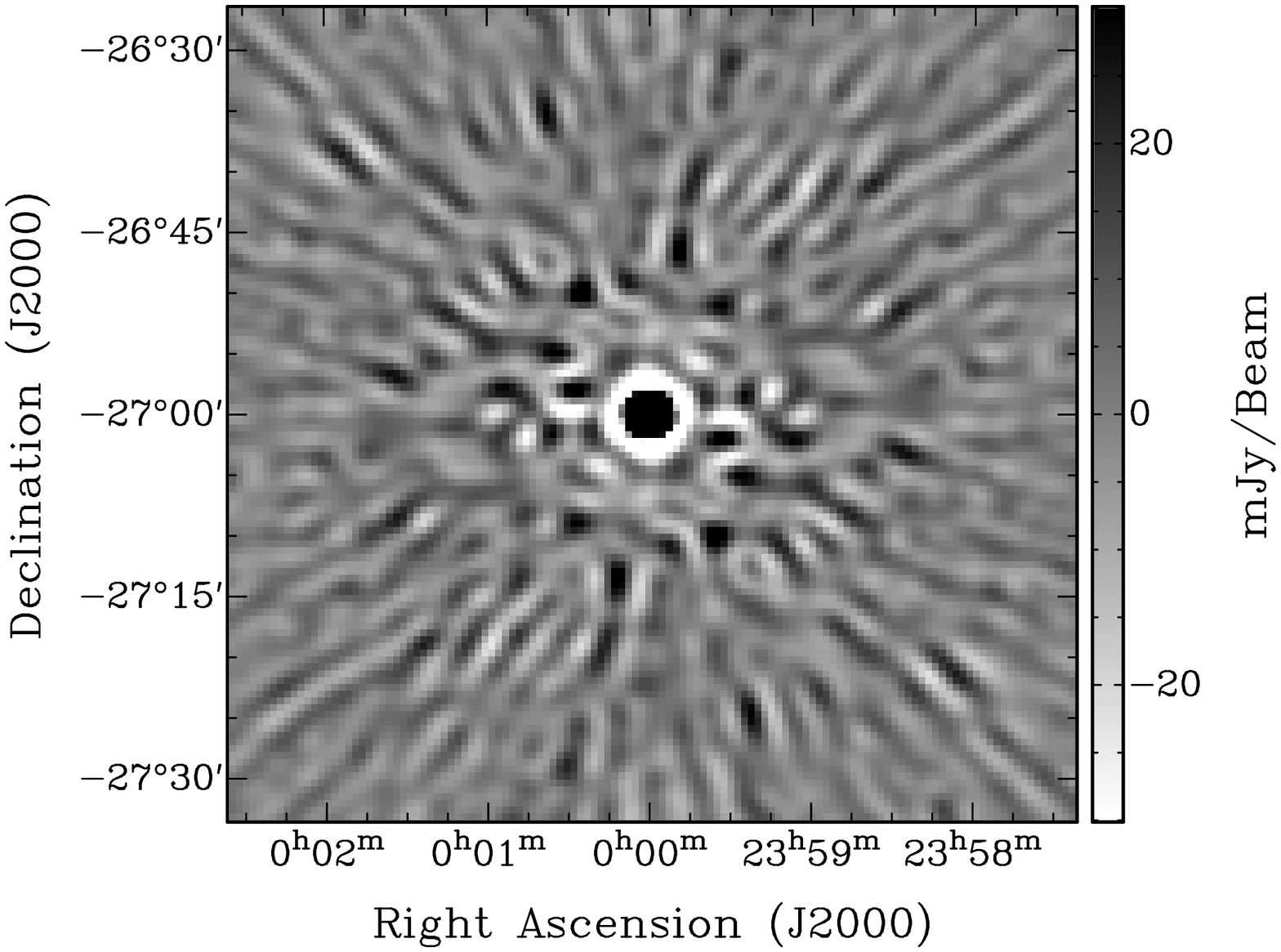}
  \includegraphics[scale=0.3,angle=270, trim=0cm 0cm -1cm 0cm]{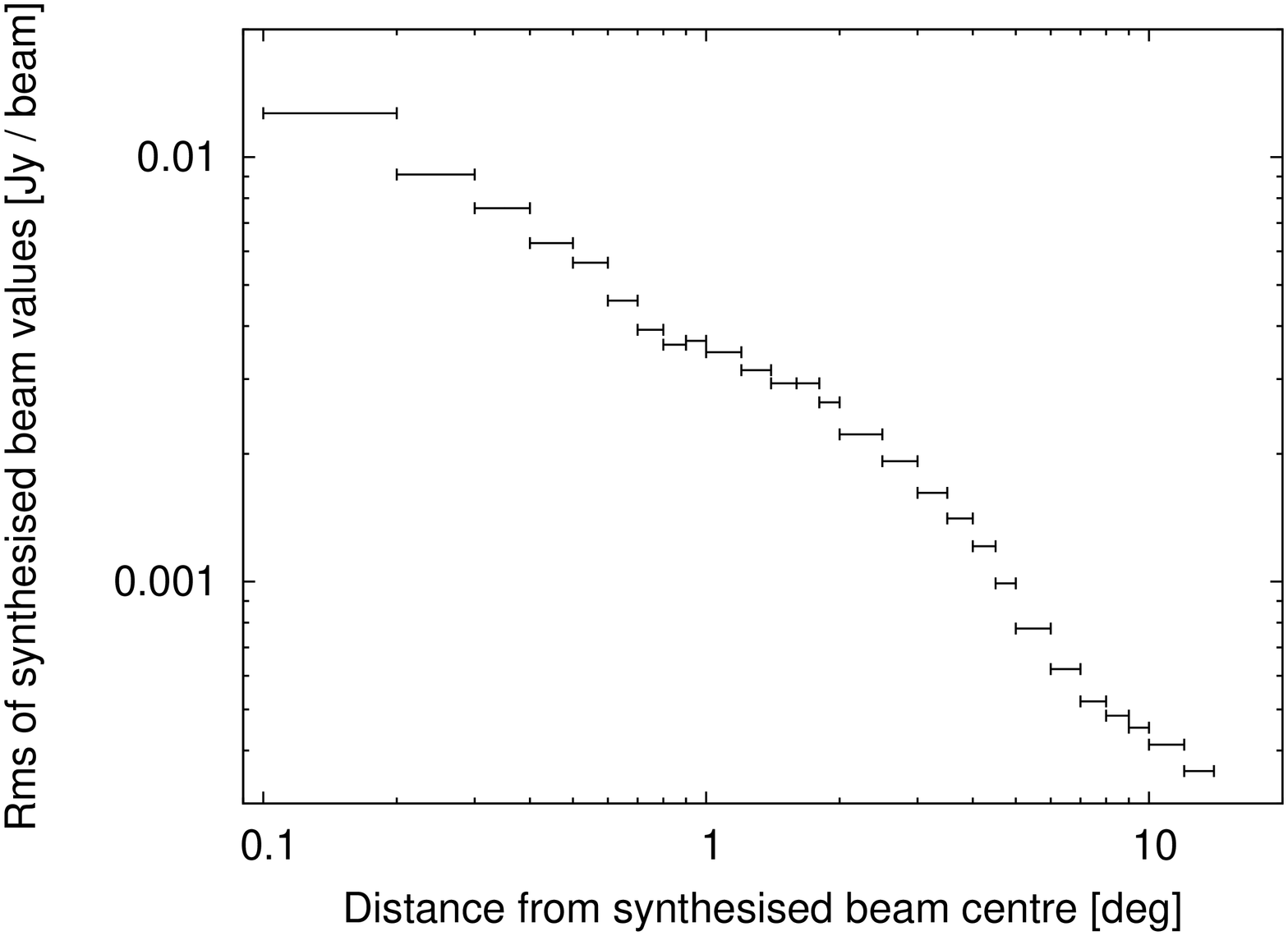}
  \caption{Top: central square degree of the MWA synthesised beam for a 2~min snapshot with a central frequency of 154~MHz and bandwidth of 30.72~MHz, using a uniform weighting scheme. The peak is 1~Jy/beam and the greyscale runs from --30 to 30~mJy/beam. The main lobe of the synthesised beam is saturated to clearly show the distant sidelobe structure. Bottom: standard deviation of the pixel values in the beam as a function of distance from the beam centre. This standard deviation was calculated in a thin annulus at the given radius.}
  \label{fig:sbeam}
 \end{center}
\end{figure}

The top panel of Fig.~\ref{fig:sbeam} shows the central square degree of the MWA synthesised beam for a 2~min snapshot with a central frequency of 154~MHz and bandwidth of 30.72~MHz, using a uniform weighting scheme. The standard deviation of the beam drops from $\approx 1.3 \times 10^{-2}$ at a distance of 10~arcmin from the beam centre to $\approx 3.5 \times 10^{-4}$ at a distance of 13.5~deg from the beam centre (i.e. at the half-power point), as shown in the bottom panel of Fig.~\ref{fig:sbeam}.

\section{MWA EoR data}\label{MWA EoR data}

\cite{offringa2016} explored the effect of foreground spectra on EoR experiments by measuring spectra with high
frequency resolution for the 586 brightest unresolved sources in the MWA EoR0 field, centred at 
J2000 $\alpha = 00^{\mathrm{h}}00^{\mathrm{m}}00^{\mathrm{s}}$, $\delta = -27^{\circ}00'00''$. The observations
used in their work were spread over 12 nights between 2013 August and 2013 October. They were
made in two frequency bands covering $139-170$~MHz and $167-198$~MHz, with a frequency resolution of 40~kHz 
and time resolution of 0.5~s.

The mean rms noise over the central 10~degrees of the Stokes $I$ image integrated over the total 59~MHz bandwidth 
was 3.6~mJy/beam after 5~h of integration. The rms noise continued to decline after 5~h of integration but not
proportionally to $1/\sqrt t$: an rms noise of 3.2~mJy/beam was reached after 45~h of integration.
The rms noise in the Stokes $V$ image continued to follow
$1/\sqrt t$, reaching a level of 0.6~mJy/beam after 45~h of integration. The Stokes $V$ image was void of sources,
except for weak sources that appeared because of instrumental leakage. The Stokes $V$ leakage was typically $0.1-1$ 
per cent of the Stokes $I$ flux density.

\begin{figure*}
\begin{center}
\includegraphics[scale=0.9,angle=270,trim=2cm 0.5cm 1cm 0cm]{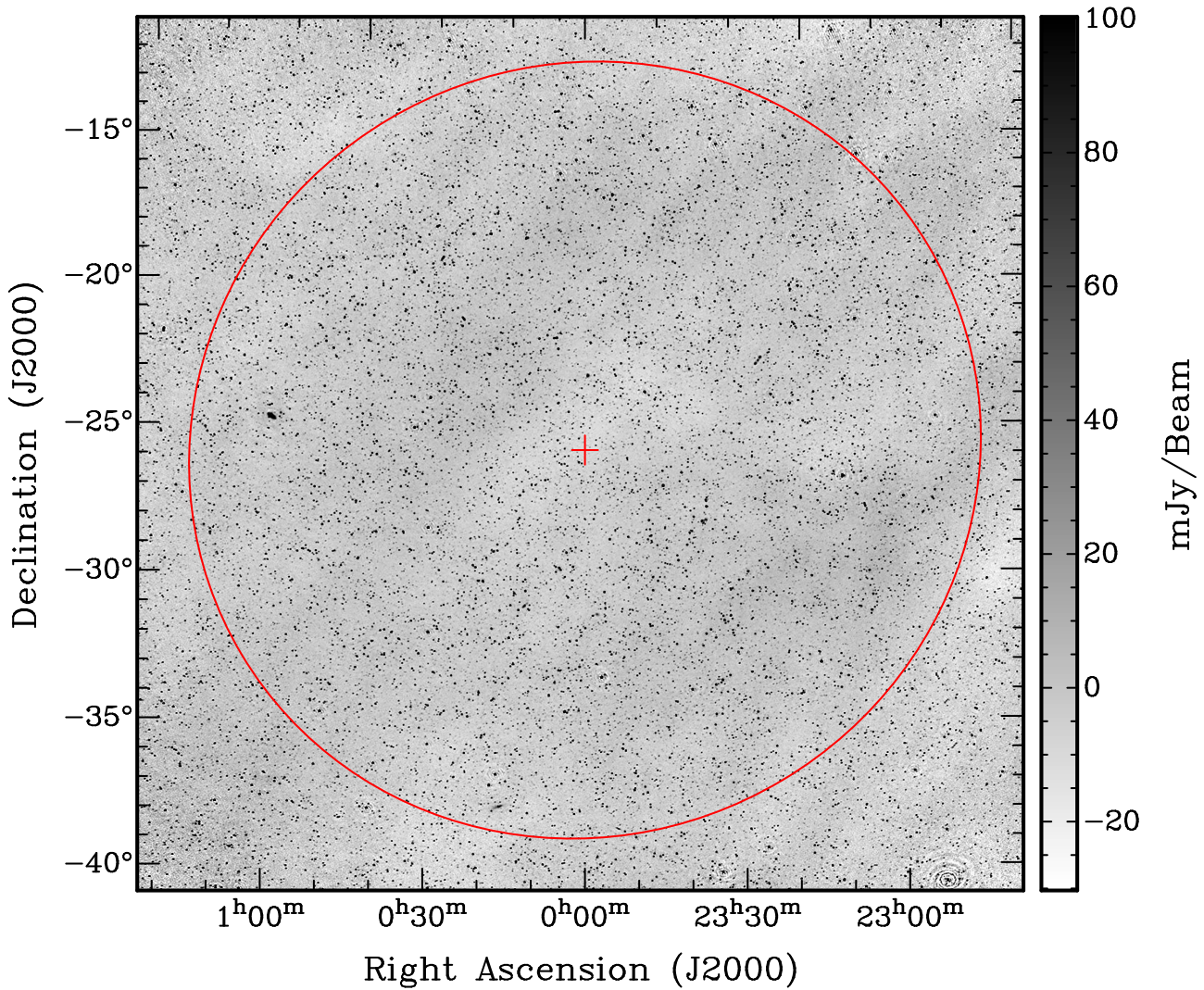}
\end{center}
\caption{Image of the EoR0 field. The red cross shows the pointing centre and the red circle the half-power
beamwidth. The greyscale is linear and runs from --30 to 100~mJy/beam.}
\label{fig:StokesI}
\end{figure*}

The image analysed in this paper was made from a 12~h subset of the low band data presented in 
\cite{offringa2016}, observed over three nights in 2013 September. The image, corrected for the primary beam, is shown in 
Fig.~\ref{fig:StokesI}. The primary beam FWHM is 27~deg
and the resolution is 2.3~arcmin. The region of the field within the primary beam FWHM covers an area of $570~\mathrm{deg}^{2}$. 

The data processing strategy is described in detail in \cite{offringa2016} and summarised here. Briefly, the \textsc{COTTER} preprocessing
pipeline \citep{offringa2015} was used to flag RFI, average the data in time to 4~s and convert the raw data to 
measurement sets; no frequency averaging was performed. 
Initial calibration was performed as a direction-independent full-polarisation self-calibration. The source
model was primarily based on the MWA Commissioning Survey \citep{hurley-walker2014} at 180~MHz and the 
Sydney University Molonglo Sky Survey \citep[SUMSS;][]{mauch2003} at 843~MHz. A few thousand sources with
apparent flux densities greater than 100~mJy, and a few complex sources, were peeled using a 
direction-dependent peeling procedure that mitigates the ionosphere. The peeled snapshots were imaged with uniform weighting
using WSClean \citep{offringa2014}. Each snapshot was CLEANed to a depth of 100~mJy/beam. The snapshots were corrected for the 
primary beam and mosaicked together. Finally, the peeled sources were added back into the mosaicked image
and restored with a Gaussian beam.

As described above, the EoR imaging process weights and adds a number of two minute snapshots. As each snapshot is short, the aperture 
plane is incompletely sampled such that all detected sources generate 
sidelobes. Each snapshot is imaged separately with the sources being deconvolved to a limit of about 4 times 
the typical rms of each snapshot. Whilst sidelobe confusion reduces as sources are extracted, there is a limit; 
eventually the sidelobe terms exceed those from real sources and any further CLEAN 
iterations will diverge rather than improve the imaging process. As a result, sidelobes from the fainter sources 
plus those from sources outside the imaged FoV remain in the 
snapshot images. The mosaicking process (weighted average of $N$ snapshots) reduces the thermal noise and 
improves the synthesised beam. However, as neither the fainter sources nor the far-field sources have been deconvolved 
in the individual snapshots, their sidelobes remain in the mosaic.

The peeling procedure more accurately characterises the synthesised beam sidelobes than CLEANing \citep[see][]{offringa2016}.
This reduction in sidelobe contamination due to peeling makes a substantial improvement to the final image.

\section{Determining the 154~MH\lowercase{z} source counts}\label{Measuring the 154 MHz source counts}

\begin{figure}
 \begin{center}
  \includegraphics[scale=0.55,angle=270, trim=4cm 5cm 2cm 0cm]{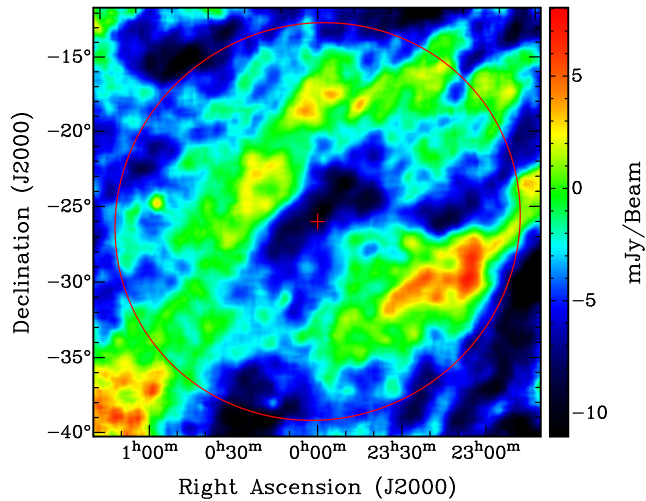}
  \caption{The background map created by running \textsc{BANE} on the EoR0 field. The red cross shows the pointing centre 
and the red circle the half-power beamwidth. The colour scale is linear and runs from --11 to 8~mJy/beam.}
  \label{fig:aegean_bkg}
 \end{center}
\end{figure}

We used the MWA EoR image described in Section~\ref{MWA EoR data} to construct a source catalogue and measure the
154~MHz counts. We first used \textsc{BANE}\footnote{https://github.com/PaulHancock/Aegean} to remove the background structure and estimate 
the noise across the image. The `box' parameter defining the angular scale on which the rms and background are evaluated
was set to 20 times the synthesised beam size. The mean rms noise over the central 10~degrees of the image is 4.6~mJy/beam.
The background map is shown in Fig.~\ref{fig:aegean_bkg}.
Large-scale fluctuations correspond to Galactic structure in that
their position is constant with frequency and they correspond to increased flux density in the Continuum HI Parkes All Sky 
Survey map \citep{calabretta2014}. The mean background within the half-power beamwidth is --2.7~mJy/beam.
We then ran the source finder \textsc{AEGEAN} \citep{hancock2012}, detecting
13,458 sources above $5 \sigma$ within 13.5~deg radius from the pointing centre.

Given that the vast majority of sources are unresolved due to the large beam size,
we used the peak flux densities to measure the counts. Source blending may significantly affect the counts because of the low 
number (25) of beams per source.
We followed a similar method to that employed by \cite{gower1966} for the 4C survey to quantify the effect 
of both source blending and incompleteness on the counts. We injected artificial point sources drawn from a source count model into the 
real image and measured the flux densities of the simulated sources using the same techniques as for
the real source list. The corrections to the counts were obtained by comparing the measured counts of 
the simulated sources with the source count model.

The source count model used was a $5^{\mathrm{th}}$ order polynomial fit to the 7C counts \citep{hales2007} at 151~MHz and the GMRT
counts by \cite{intema2011}, \cite{ghosh2012} and \cite{williams2013} at 153~MHz.
As these counts are measured at very similar frequencies, we neglect any correction to transpose them to 154~MHz as the effect will be inconsequential; henceforth, when referring to the 7C and GMRT counts, we consider them to be measured at 154~MHz.
A total of 32,120 sources with flux densities ranging between 6~mJy and 100~Jy 
were injected into the region of the field within the primary beam FWHM using the \textsc{miriad} task \textsc{imgen}.
The positions of the simulated sources were chosen randomly; to avoid the simulated sources affecting each other, 
a simulated source was not permitted to lie within 10~arcmin from any other and only 3,212 sources were injected into the image at a time.

In cases where a simulated source was found to lie too close to a real source to be 
detected separately, the simulated source was considered to be detected if the recovered source position was closer
to the simulated rather than the real source position. The Monte Carlo analysis therefore accounts
for sources which are omitted from the source finding process through being too close to a brighter source.

\begin{figure}
 \begin{center}
  \includegraphics[scale=0.33,angle=270]{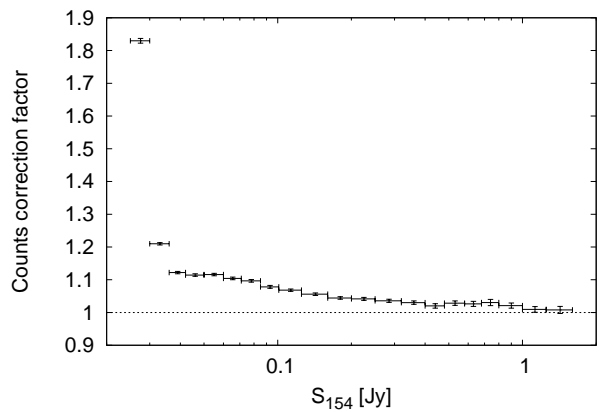}
  \caption{Top: Source count correction factor as a function of flux density, accounting for both source confusion and incompleteness.}
  \label{fig:corr_counts}
 \end{center}
\end{figure}

These simulations were repeated 100 times to improve statistics.
Fig.~\ref{fig:corr_counts} shows the mean correction factor to be applied to the counts in each flux density bin. 
The error bars are standard errors of the mean. The gradual increase
in the correction factor between $\approx 1$~Jy and $\approx 40$~mJy is due to source confusion; the effect of confusion is to 
steepen the slope of the counts. The sharp increase in the correction factor below $\approx 40$~mJy is due to incompleteness.

\begin{figure*}
\begin{center}
\includegraphics[scale=0.7,angle=270, trim=0cm 0cm 0cm 0cm]{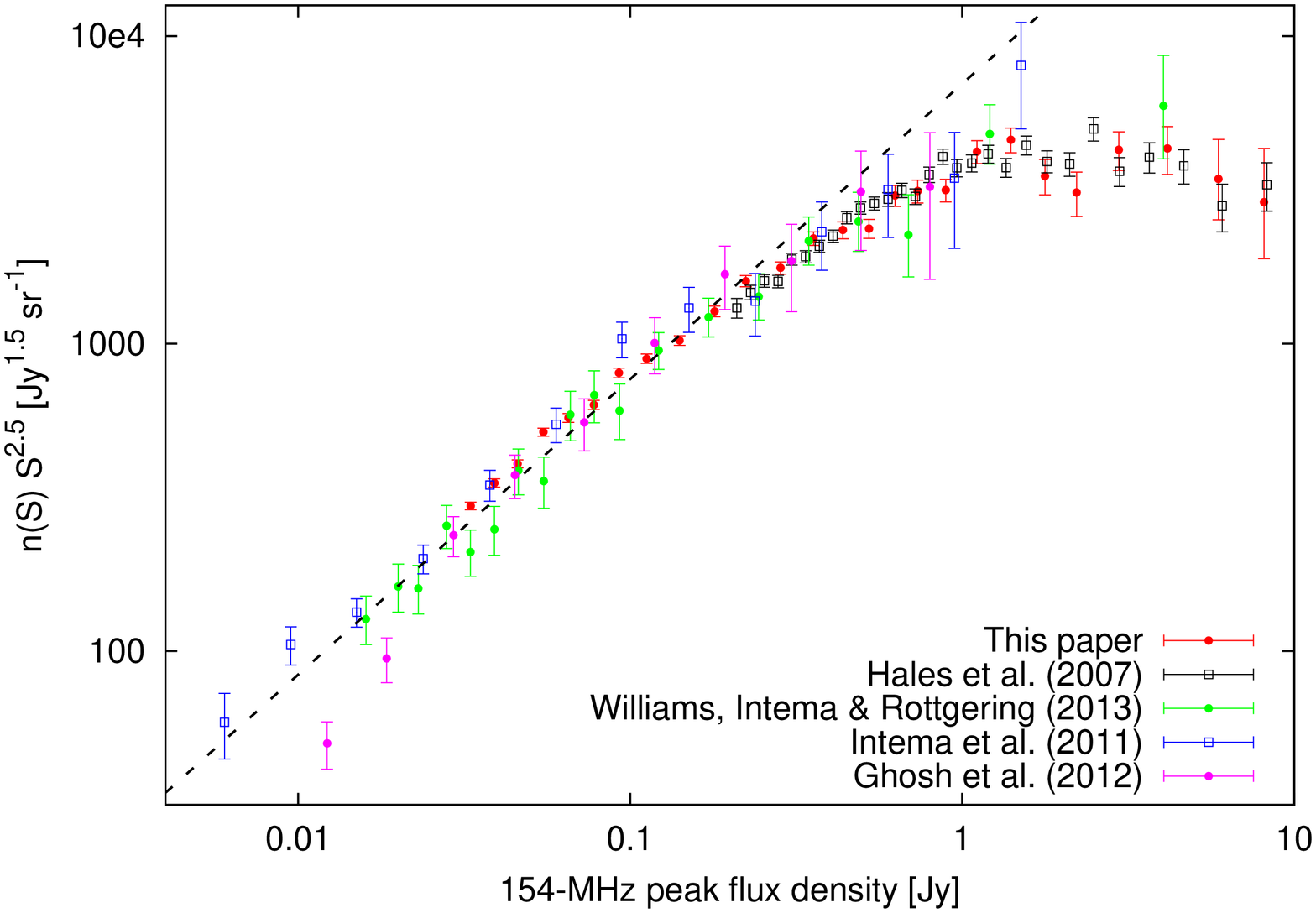}
\end{center}
\caption {Euclidean normalised ($S^{2.5} \frac{dN}{dS}$) differential counts at 154~MHz.
Red circles: this paper; black squares: \citet{hales2007}; 
green circles: \citet{williams2013}; blue squares: \citet{intema2011}; purple circles: 
\citet{ghosh2012}. The dashed diagonal line shows a weighted least-squares power-law 
fit ($\frac{dN}{dS} = 6998~S^{-1.54} \, \mathrm{Jy}^{-1} \mathrm{sr}^{-1}$) 
to the GMRT data from \citet{williams2013}, \citet{intema2011} and \citet{ghosh2012} below 400~mJy.}
\label{fig:counts_MWA}
\end{figure*}

Our MWA differential source count data corrected for
incompleteness and source blending are provided in Table~\ref{tab:mwa_source_counts}.
Fig.~\ref{fig:counts_MWA} shows the MWA counts compared with the 7C and GMRT counts.
We find that the MWA counts are in excellent agreement with the other counts
despite the lower MWA resolution: in comparison, the 7C survey has a resolution of 70~arcsec and the GMRT observations
have resolutions ranging between 20 and 25~arcsec. This indicates that the flux density scale of the MWA image, based on the
MWA Commissioning Survey \citep{hurley-walker2014}, is fully consistent with the 7C survey and the deeper GMRT data.
The MWA counts are by far the most precise in the flux density range 30--200~mJy as a result of the large area of sky covered ($570~\mathrm{deg}^{2}$). 
The surveys used to derive the GMRT counts cover areas ranging between 6.6 and $30~\mathrm{deg}^{2}$.

\begin{table*}
\centering
\caption{Differential counts at 154~MHz in EoR0 from the MWA image. The counts are
corrected for incompleteness and source blending as described in the text. The bin centre 
corresponds to the mean flux density of all sources in each bin.
In the two highest flux density bins which contain less than 20 sources each, we use approximate formulae for confidence
limits based on Poisson statistics by \citet{gehrels1986}. In the remaining bins, the Poisson error on the
number of sources is approximated as the square root of the number of sources.}
\label{tab:mwa_source_counts}
\begin{tabular}{@{} c c c c c c} 
\hline
\multicolumn{1}{c}{Bin start}
&\multicolumn{1}{c}{Bin end}
&\multicolumn{1}{c}{Bin centre}
&\multicolumn{1}{c}{Number of}
&\multicolumn{1}{c}{Correction}
&Euclidean normalised\\
\multicolumn{1}{c}{$S$ (Jy)}
&\multicolumn{1}{c}{$S$ (Jy)}
&\multicolumn{1}{c}{$S$ (Jy)}
&\multicolumn{1}{c}{sources}
&\multicolumn{1}{c}{factor}
&\multicolumn{1}{c}{counts ($\mathrm{Jy}^{3/2} \mathrm{sr}^{-1}$)} \\
\hline
    7.00 &    10.00 &     8.12 &          8 & - & $      2884^{1428}_{-997}$ \\
    5.00 &     7.00 &     5.92 &         14 & - & $      3431^{1186}_{-905}$ \\
    3.50 &     5.00 &     4.15 &         32 & - & $      4315 \pm        763 $ \\
    2.50 &     3.50 &     2.96 &         49 & - & $      4269 \pm        610 $ \\
    2.00 &     2.50 &     2.21 &         37 & - & $      3101 \pm        510 $ \\
    1.60 &     2.00 &     1.78 &         58 & - & $      3506 \pm        460 $ \\
    1.25 &     1.60 &     1.40 &        119 & $     1.008 \pm      0.011 $ & $      4599 \pm        424 $ \\
    1.00 &     1.25 &     1.11 &        140 & $     1.010 \pm      0.009 $ & $      4211 \pm        358 $ \\
    0.800 &     1.000 &     0.892 &        143 & $     1.021 \pm      0.008 $ & $      3157 \pm        265 $ \\
    0.680 &     0.800 &     0.735 &        137 & $     1.031 \pm      0.010 $ & $      3135 \pm        269 $ \\
    0.580 &     0.680 &     0.628 &        164 & $     1.026 \pm      0.008 $ & $      3032 \pm        238 $ \\
    0.480 &     0.580 &     0.524 &        201 & $     1.029 \pm      0.007 $ & $      2366 \pm        168 $ \\
    0.400 &     0.480 &     0.437 &        252 & $     1.020 \pm      0.007 $ & $      2342 \pm        148 $ \\
    0.320 &     0.400 &     0.357 &        390 & $     1.030 \pm      0.005 $ & $      2198 \pm        112 $ \\
    0.250 &     0.320 &     0.284 &        482 & $     1.036 \pm      0.005 $ & $      1764 \pm         81 $ \\
    0.200 &     0.250 &     0.223 &        567 & $     1.042 \pm      0.004 $ & $      1597 \pm         67 $ \\
    0.160 &     0.200 &     0.180 &        621 & $     1.044 \pm      0.004 $ & $      1275 \pm         51 $ \\
    0.125 &     0.160 &     0.141 &        793 & $     1.056 \pm      0.004 $ & $      1024 \pm         37 $ \\
    0.101 &     0.125 &     0.112 &        829 & $     1.068 \pm      0.004 $ & $       893 \pm         31 $ \\
    0.085 &     0.101 &     0.0925 &        795 & $     1.078 \pm      0.004 $ & $       803 \pm         29 $ \\
    0.071 &     0.085 &     0.0777 &        833 & $     1.097 \pm      0.004 $ & $       632 \pm         22 $ \\
    0.060 &     0.071 &     0.0652 &        916 & $     1.104 \pm      0.004 $ & $       574 \pm         19 $ \\
    0.050 &     0.060 &     0.0548 &       1141 & $     1.116 \pm      0.004 $ & $       516 \pm         15 $ \\
    0.042 &     0.050 &     0.0458 &       1129 & $     1.114 \pm      0.004 $ & $       407 \pm         12 $ \\
    0.036 &     0.042 &     0.0390 &       1090 & $     1.122 \pm      0.004 $ & $       352 \pm         11 $ \\
    0.030 &     0.036 &     0.0331 &       1284 & $     1.210 \pm      0.004 $ & $       297 \pm          8 $ \\
\hline
\end{tabular}
\end{table*}

\section{Quantifying the classical and sidelobe confusion noise}\label{Quantifying the classical and sidelobe confusion noise}

We use the method of probability of deflection, or $P(D)$ analysis, to quantify the classical and sidelobe confusion in the MWA EoR image.
We also investigate the effect of image artefacts caused by calibration errors on the pixel statistics, and estimate the degree of bandwidth 
and time-average smearing in the image.

The deflection $D$ at any pixel in the image is the intensity in units of mJy/beam. We assume that the observed $P(D)$ distribution is given by
\begin{eqnarray}
\label{equation:P(D)}
P_{\mathrm{o}}(D) = P_{\mathrm{n}}(D) \ast P_{\mathrm{c}}(D) \ast P_{\mathrm{s}}(D)
\mathrm{,}
\end{eqnarray}
where `$\ast$' represents convolution, $P_{\mathrm{n}}(D)$ is the $P(D)$ distribution resulting from the Gaussian system noise, $P_{\mathrm{c}}(D)$ is the $P(D)$ distribution
from all sources present in the image given the synthesised beam size and $P_{\mathrm{s}}(D)$ is the
$P(D)$ distribution from residual sidelobes. We take $P_{\mathrm{s}}(D)$ to include image artefacts due to calibration errors and smearing effects. 

Below 400~mJy, the Euclidean normalised differential counts at 154~MHz from \cite{williams2013},
\cite{intema2011} and \cite{ghosh2012} are well represented by a power law of the form $\frac{dN}{dS} = kS^{-\gamma} \, \mathrm{Jy}^{-1} \mathrm{sr}^{-1}$, 
with $k = 6998$ and $\gamma = 1.54$ (see dashed diagonal line in Fig.~\ref{fig:counts_MWA}). In these data we see that the 
154~MHz differential source counts continue to decline at $S_{154} \lesssim 10$~mJy. Any flattening towards low flux densities, 
seen at higher frequencies, has not yet been detected. 

We derived $P_{\mathrm{c}}(D)$ for this source count model fit as follows: we simulated a 9.63 by 9.63~deg noise-free image 
containing point sources at random positions, randomly assigning their flux densities from the power-law fit.
A total of 51,887 sources with flux densities ranging between 0.1 and 400~mJy were injected into the image.
The simulated point sources were convolved with the Gaussian restoring beam; 
we did not attempt to model the sidelobe confusion. Although our source count model fit is valid between 6 and 400~mJy,
we assumed no change in the source count slope below 6~mJy. In Section~\ref{Extending the observed source count}, 
we explore the effect of a flattening in the counts below 6~mJy on the classical confusion noise. 
We obtained $P_{\mathrm{c}}(D)$ from the distribution of pixel values in the simulated image.

\begin{figure}
\begin{center}
\includegraphics[scale=0.33,angle=270]{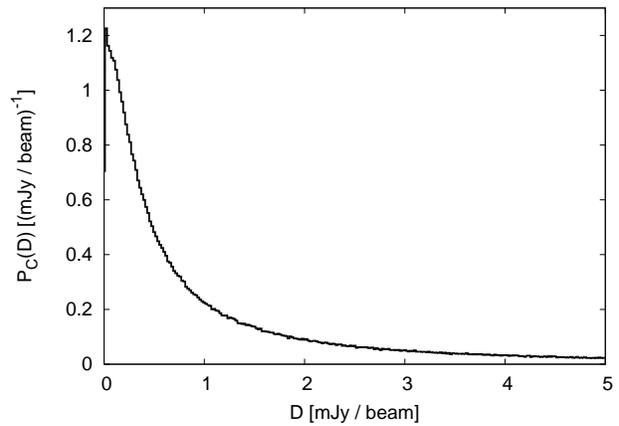}
\end{center}
\caption {$P_{\mathrm{c}}(D)$ assuming the extrapolated source count fit of Fig.~\ref{fig:counts_MWA} 
and a Gaussian beam of size 2.31~arcmin FWHM, calculated as discussed in Section~\ref{Quantifying the classical and sidelobe confusion noise}.}
\label{fig:source_dist}
\end{figure}

Fig.~\ref{fig:source_dist} shows the derived source $P(D)$ distribution. The width of this distribution is 1.7~mJy/beam, as
measured by dividing the interquartile range by 1.349, i.e. the rms for a 
Gaussian distribution. As noted by \cite{zwart2015}, source $P(D)$ distributions are usually highly skewed
and very non-Gaussian. Although we have quoted the core width of the distribution, 
we caution that a single descriptor is unsuitable. Of course, the advantage of a $P(D)$ analysis is
that it accounts for the exact shape of the distribution. 

\begin{figure}
\begin{center}
\includegraphics[scale=0.33,angle=270]{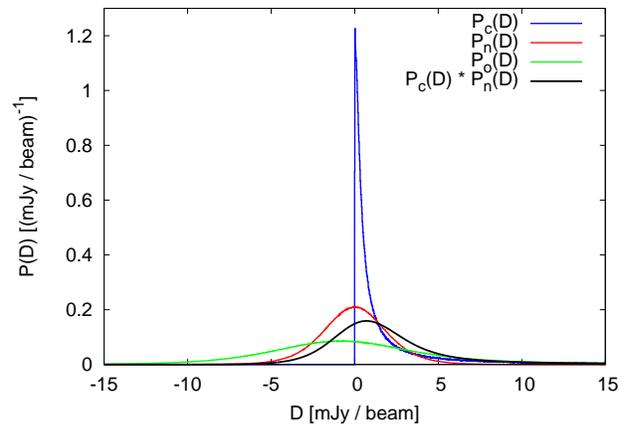}
\end{center}
\caption {Source $P(D)$ distribution as in Fig.~\ref{fig:source_dist} (blue), pixel distribution in the Stokes $V$ image, 
representing the system noise distribution (red), pixel distribution in the Stokes $I$ image (green) and 
convolution of the source $P(D)$ distribution with the system noise distribution (black).}
\label{fig:modelA}
\end{figure}

Fig.~\ref{fig:modelA} shows $P_{\mathrm{c}}(D)$, $P_{\mathrm{n}}(D)$ as represented by the pixel distribution in the Stokes $V$ image,
$P_{\mathrm{o}}(D)$ as represented by the pixel distribution in the Stokes $I$ image and
$P_{\mathrm{c}}(D) \ast P_{\mathrm{n}}(D)$. If the source count model is correct and the sidelobe confusion is negligible, 
$P_{\mathrm{c}}(D) \ast P_{\mathrm{n}}(D)$ would agree with $P_{\mathrm{o}}(D)$.

\begin{figure}
 \begin{center}
  \includegraphics[scale=0.33,angle=270]{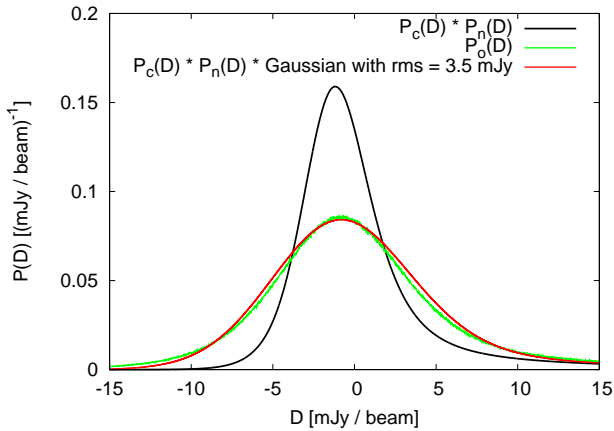}
  \caption{The green curve shows $P_{\mathrm{o}}(D)$, the black curve shows $P_{\mathrm{c}}(D) \ast P_{\mathrm{n}}(D)$, shifted by 1.88~mJy/beam
to the left to remove the offset in the $x$-direction with respect to $P_{\mathrm{o}}(D)$, and the red curve shows the black curve 
convolved with a Gaussian with $\mathrm{rms} = 3.5$~mJy/beam; this Gaussian is taken to represent $P_{\mathrm{s}}(D)$.}
  \label{fig:modelA_sidelobes}
 \end{center}
\end{figure}

\begin{figure}
 \begin{center}
  \includegraphics[scale=0.3,angle=270]{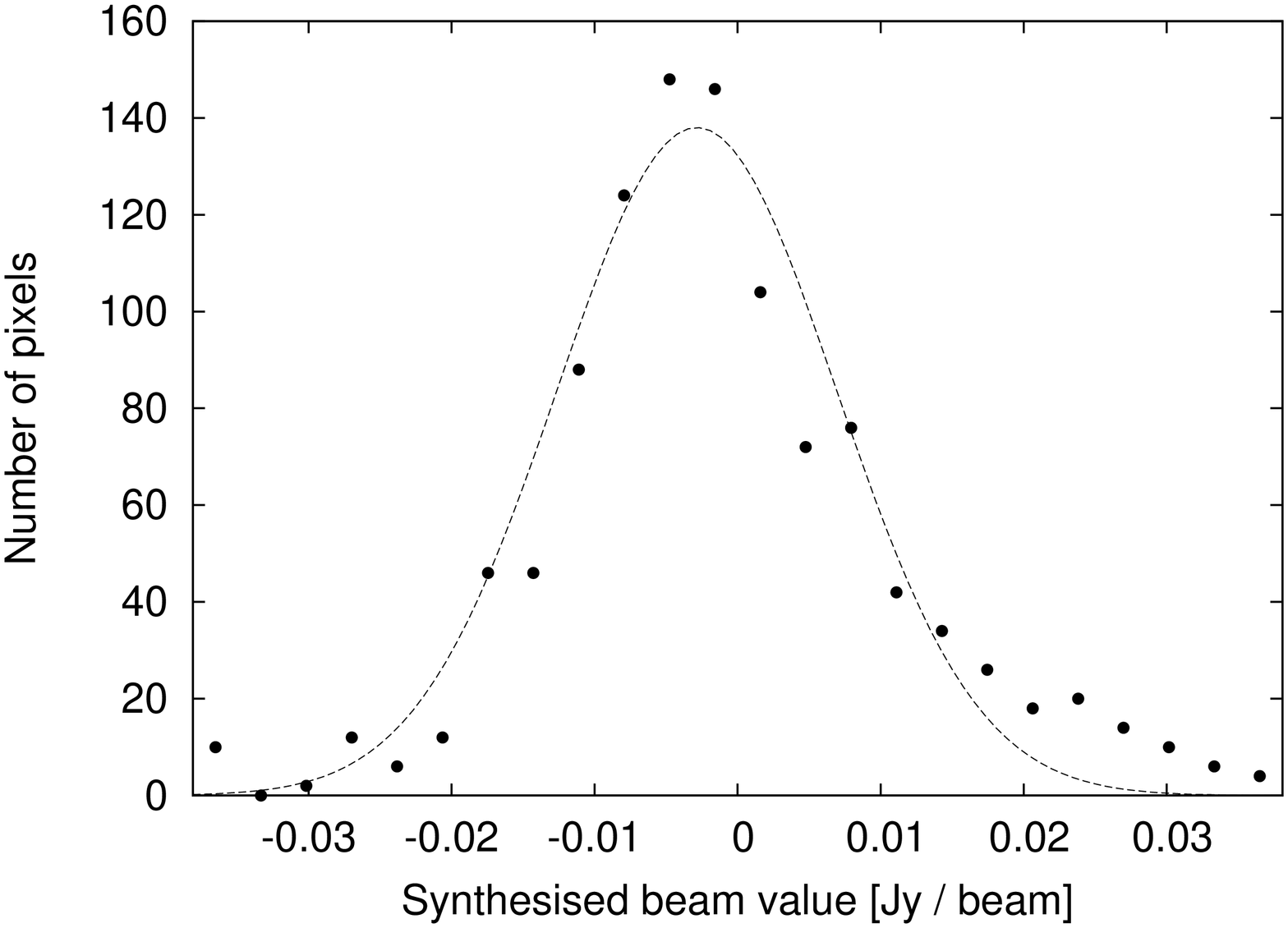}
  \includegraphics[scale=0.3,angle=270]{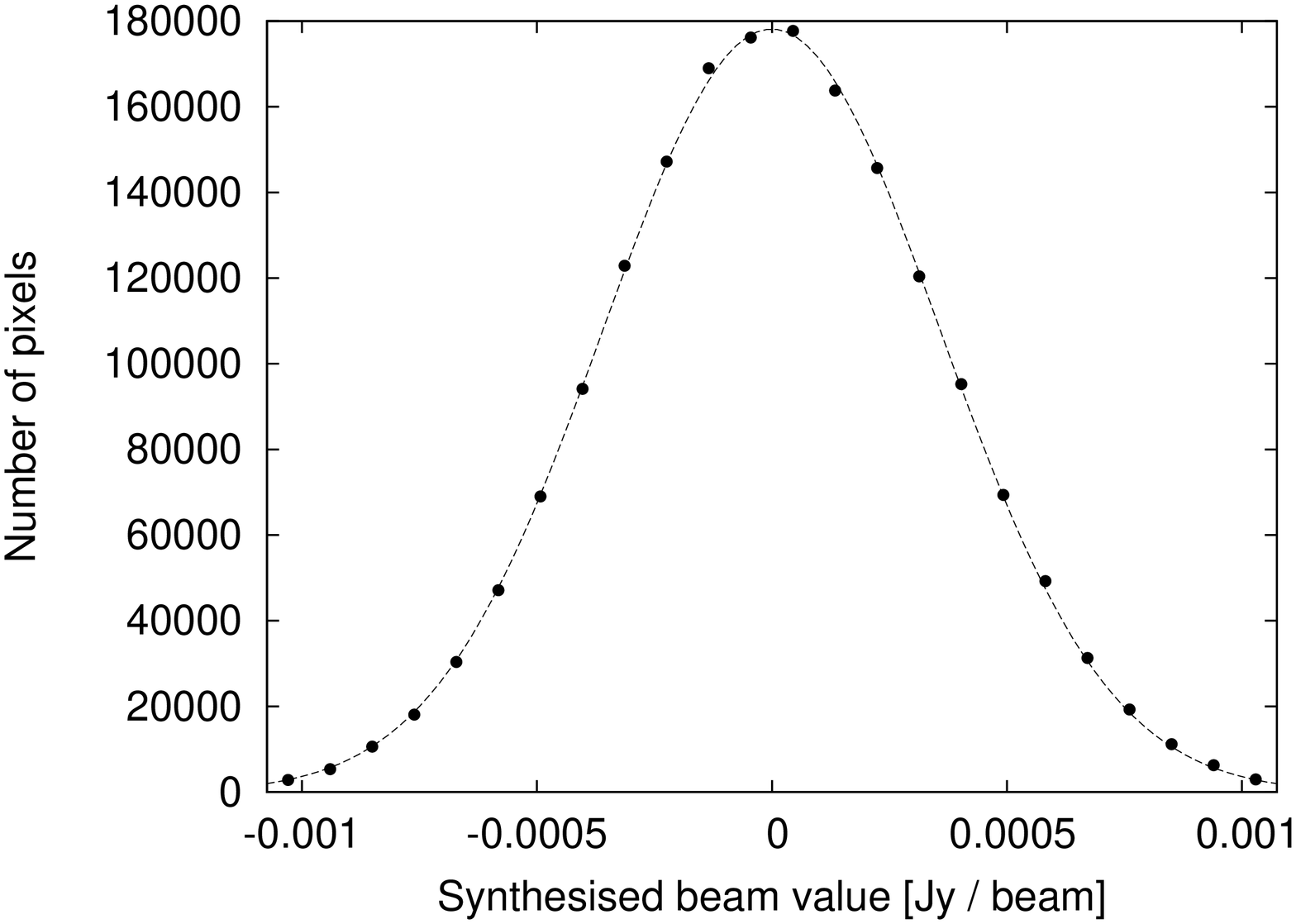}
  \caption{Top: the circles show the distribution of synthesised beam pixel values in an annulus centred on the beam with inner and outer radii of 0.1 and 0.2~deg, respectively. The dashed line is a least-squares Gaussian fit to the data points. Bottom: distribution of synthesised beam pixel values in an annulus centred on the beam with inner and outer radii of 12 and 14~deg, respectively.}
  \label{fig:sbeam2}
 \end{center}
\end{figure}

However, as can be seen in Fig.~\ref{fig:modelA} this is not the case, and requires further interpretation:
the image zero-point $D=0$ can be treated as a free parameter when comparing the observed $P(D)$ distribution with 
models of the source $P(D)$ distribution (the background was subtracted from the real image, and in any case,
interferometers have no sensitivity to large-scale emission). 
Fig.~\ref{fig:modelA_sidelobes} compares $P_{\mathrm{c}}(D) \ast P_{\mathrm{n}}(D)$ with $P_{\mathrm{o}}(D)$ after 
removing an offset of 1.88~mJy/beam in the x-direction between the two curves, where there remains very poor agreement.
Given the excellent $uv$ coverage of the MWA and the huge number of sources present in the FoV, it is reasonable to expect the sidelobe 
confusion noise to be roughly Gaussian. Indeed, we examined the distribution of pixel values in 27 thin annuli centred on the synthesised 
beam with radii ranging between 0.15 and 13~deg. In each case, the distribution of pixel values was found to be approximately 
Gaussian. Fig.~\ref{fig:sbeam2} shows the distribution at the two extremes.

We subsequently convolved $P_{\mathrm{c}}(D) \ast P_{\mathrm{n}}(D)$ with a Gaussian with rms = 3.5~mJy/beam, obtaining the red curve in Fig.~\ref{fig:modelA_sidelobes}, which is very close to $P_{\mathrm{o}}(D)$. We therefore conclude that the sidelobe confusion noise is $\approx 3.5$~mJy/beam, on the assumption that the extrapolated ($S < 6$~mJy) source count model remains valid.

\subsection{Calibration artefacts}\label{Calibration artefacts}

\begin{figure}
 \begin{center}
  \includegraphics[scale=0.5,angle=270, trim=4cm 2cm 2cm 0cm]{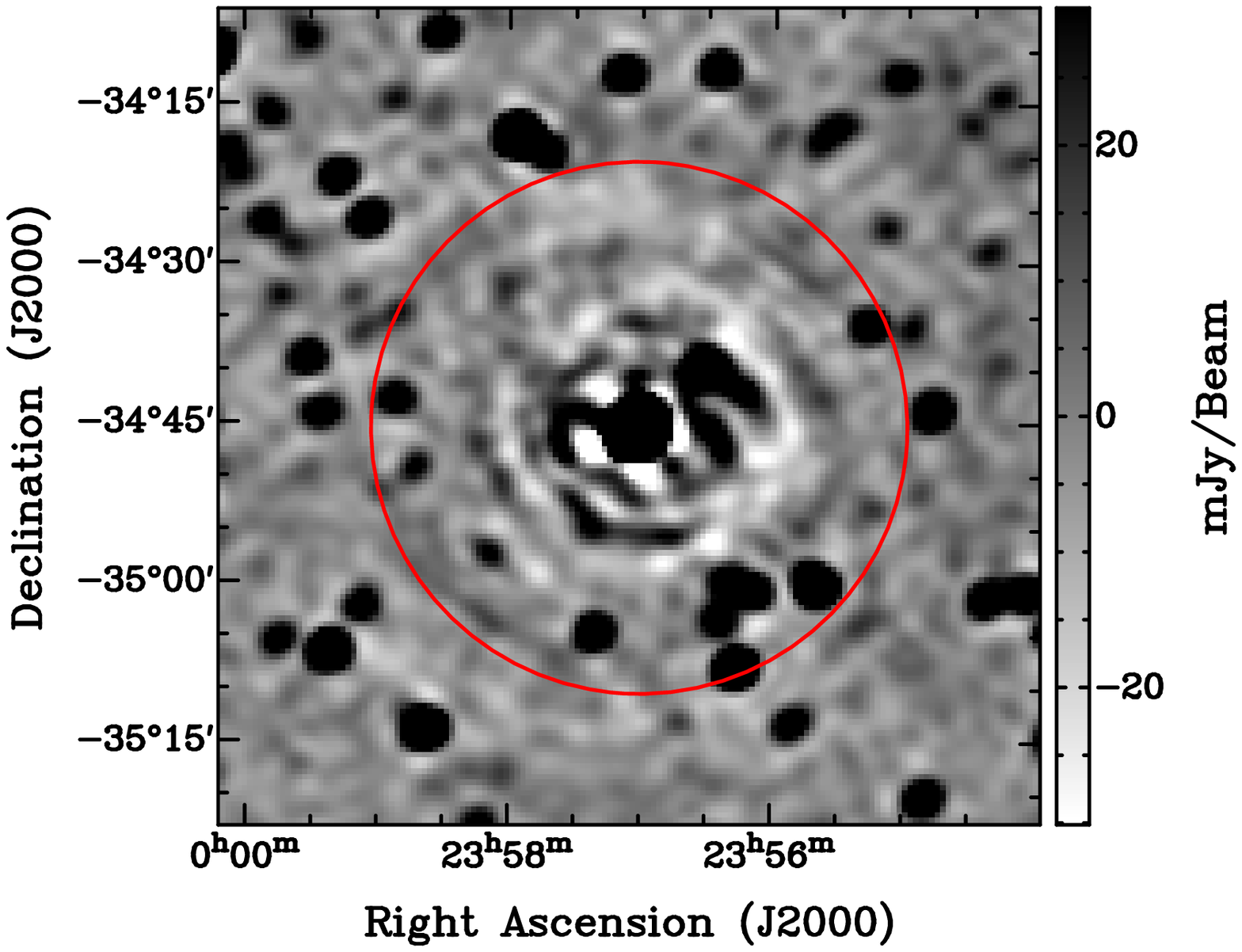}
  \caption{Section of the EoR field centred on a 24.3~Jy source. Small errors in the amplitude and phase calibration of the visibility data lead to artefacts in the image. The subsequent exclusion region around this source is shown as a red circle.}
  \label{fig:excl_zone}
 \end{center}
\end{figure}

There is an increased level of noise around the brightest sources in the field due to calibration errors
(see Fig.~\ref{fig:excl_zone}). The rms in the vicinity of sources above 5~Jy, lying within the half-power beamwidth, is typically 0.1 per cent
of the source's peak flux density. For a source of peak flux density $S_{\mathrm{pk}}$, the noise was found to be elevated within a 
distance $R$ from the source, where

\begin{equation}
R = 5 \left(S_{\mathrm{pk}}/\mathrm{Jy}\right)^{1/2}~\mathrm{arcmin} \mathrm{.}
\end{equation}

We repeated the analysis described in Section~\ref{Quantifying the classical and sidelobe confusion noise} after masking all 
pixels in the Stokes $I$ image lying within distance $R$ from sources with $S_{\mathrm{pk}} > 5.0$~Jy. The fraction of pixels 
excised from the map was 1.1 per cent. 
We obtained $\sigma_{\mathrm{s}} = 3.4$~mJy/beam, which is very close to our estimate of $\sigma_{\mathrm{s}}$ (3.5~mJy/beam) in the original 
image, indicating that calibration artefacts have a negligible effect on $\sigma_{\mathrm{s}}$.

\subsection{Wide-field imaging effects}\label{Wide-field imaging effects}

Assuming a Gaussian beam and rectangular bandpass, bandwidth smearing causes the peak flux density of a point source in 
an individual snapshot to be multiplied by

\begin{equation}
\alpha = \left[1 + \frac{2 {\rm ln}2}{3}\left(\frac{\Delta\nu_{\rm eff}}{\nu}\frac{d}{\theta}\right)\right]^{-\frac{1}{2}} \leq 1\ ,
\end{equation}

where $\Delta\nu_{\rm eff}$ is the effective channel bandwidth, $\nu$ the central frequency, $d$ the radial distance 
from the phase centre and $\theta$ the synthesised beam FWHM \citep{condon1998}. The width of the source in the radial direction
is divided by $\alpha$. For $\delta \nu = 40$~kHz, $\nu = 154$~MHz, $\theta = 2.31$~arcmin and $d = 13.5$~deg, i.e. 
at the half-power point, $\alpha = 0.980$. Given that each snapshot was weighted by the square
of its primary beam response in the mosaicking process, we have established that $0.980 < \alpha < 1$ 
in the final mosaicked image within 13.5~deg from the image centre. The bandwidth smearing effect is small and cannot contribute
significantly to $P_{\mathrm{s}}(D)$.

For uniform circular $uv$ coverage, time-average smearing causes the peak flux density of a point source near the 
North or South Celestial Pole to be multiplied by

\begin{equation}
\beta = 1 - 1.08 \times 10^{-9} \left(\frac{d}{\theta}\right)^{2} \tau^{2} \leq 1\ ,
\end{equation}

where $\tau$ is the averaging time \citep{bridle1999}. The width of the source in the azimuthal direction is 
divided by $\beta$. For $\tau = 4$~s, $\theta = 2.31$~arcmin, and $d = 13.5$~deg, $\beta = 0.998$. 
The time-average smearing effect is even smaller than the bandwidth smearing effect.

\section{Extending the observed 154~MH\lowercase{z} source count}\label{Extending the observed source count}

\begin{figure}
 \begin{center}
\includegraphics[scale=0.33,angle=270, trim=0cm 0cm 0cm 0cm]{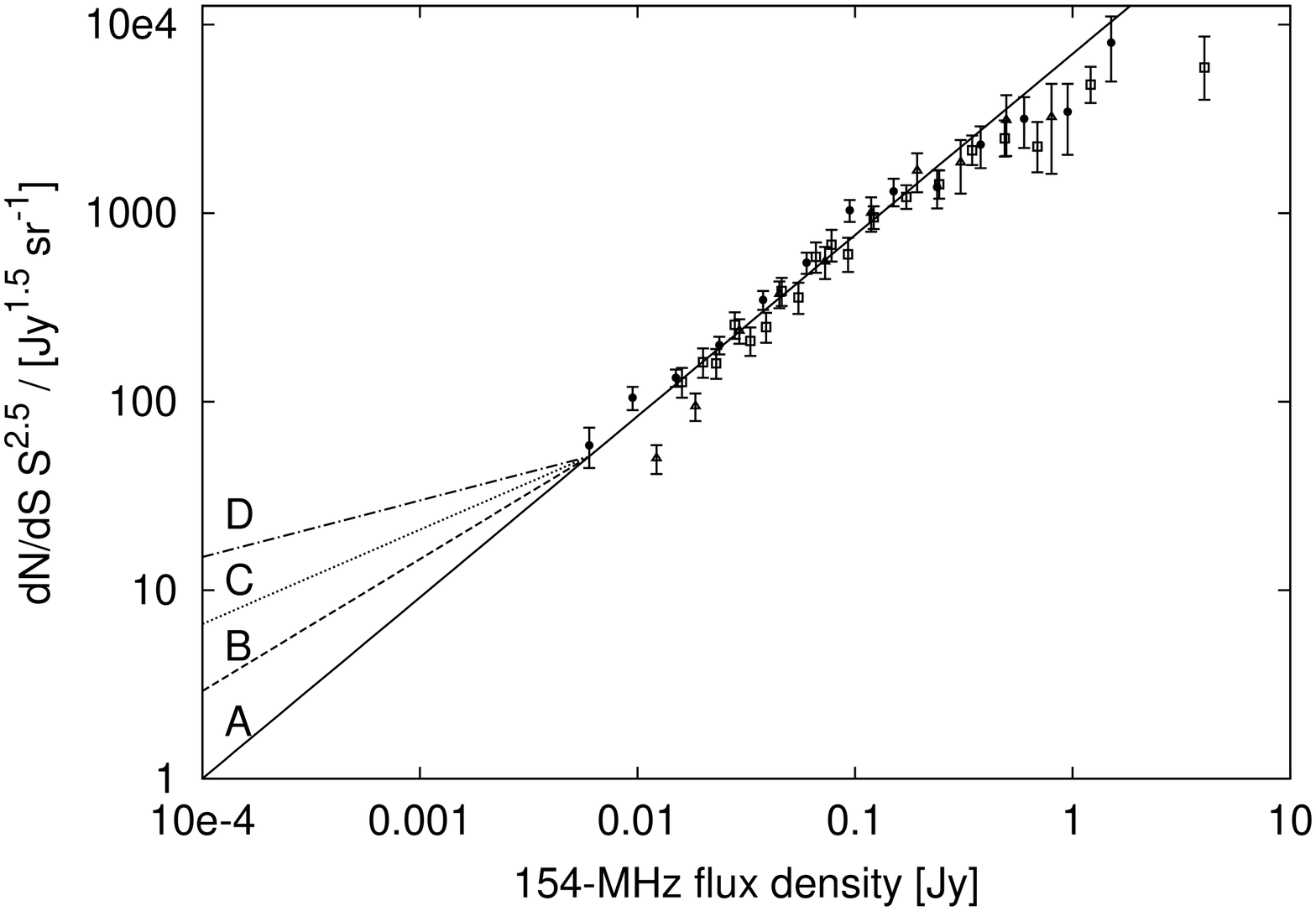}
  \includegraphics[scale=0.33,angle=270]{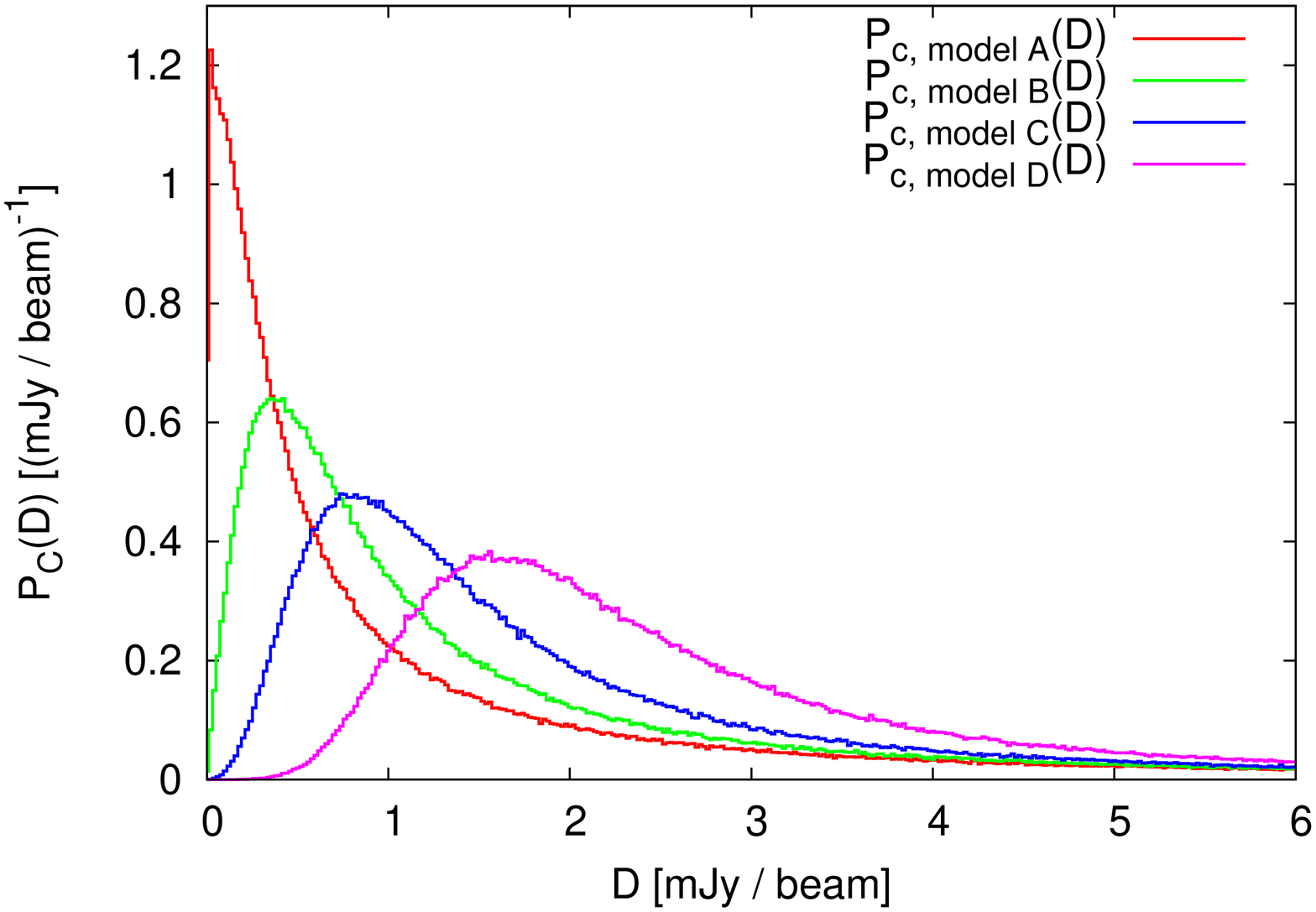}
  \includegraphics[scale=0.33,angle=270]{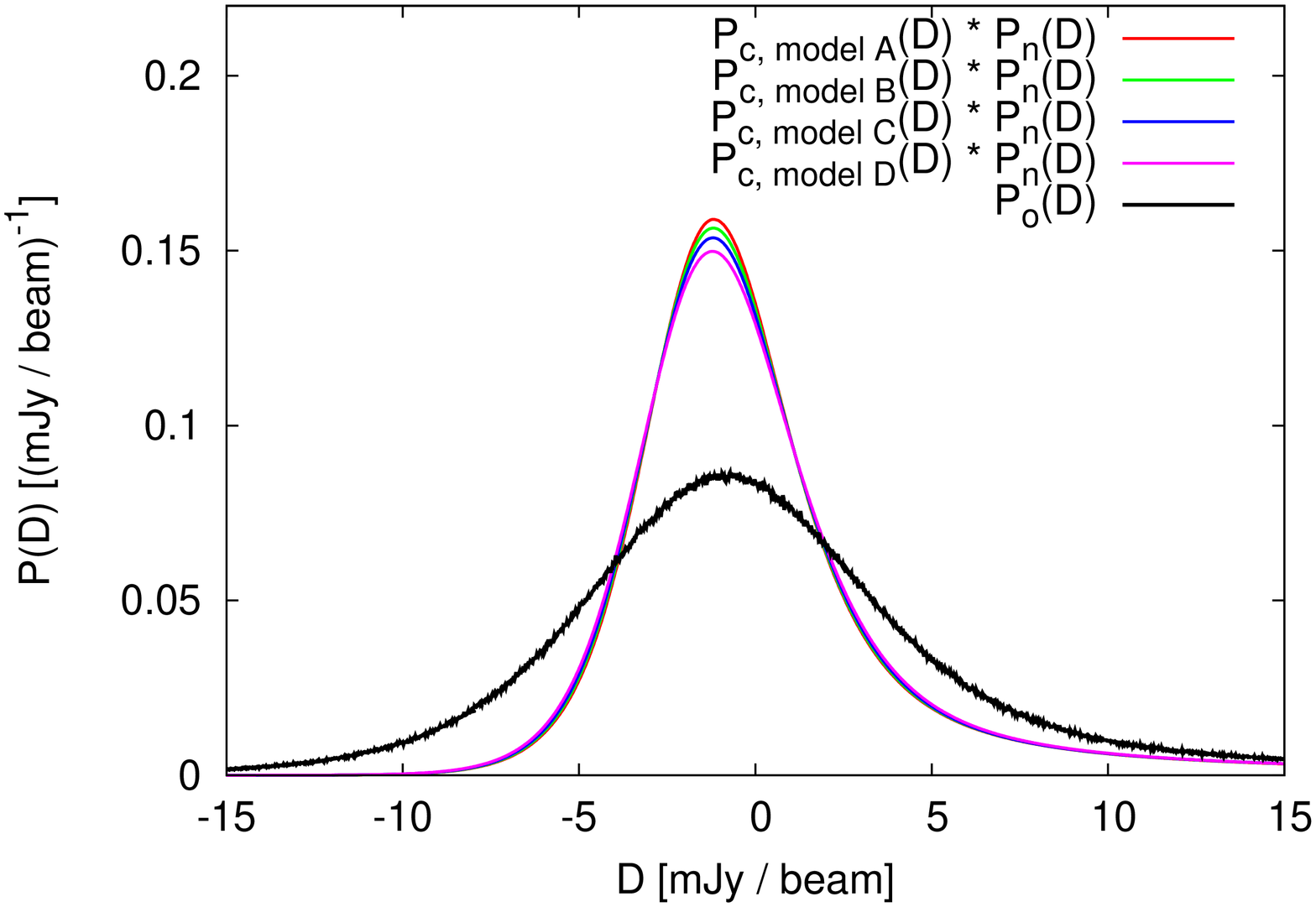}
  \caption{Top: The squares, circles and triangles show the 154~MHz counts by \citet{williams2013}, \citet{intema2011}
and \citet{ghosh2012}, respectively. The solid, dashed, dotted and dot-dashed lines 
show source count models A, B, C and D, respectively.
Middle: $P_{\mathrm{c}}(D)$ distributions corresponding to source count models A--D, using a Gaussian beam with 
$\mathrm{FWHM} = 2.31$~arcmin. The rms values of these distributions are quoted in Table~\ref{tab:scount_model_par}.
Bottom: observed $P_{\mathrm{o}}(D)$ distribution (black), $P_{\mathrm{c}}(D)$ distributions for models A (red), B (green), C (blue) and 
D (purple), convolved with the noise $P_{\mathrm{n}}(D)$ distribution and shifted to the left to remove offsets in the
x-direction with respect to $P_{\mathrm{o}}(D)$.}
  \label{fig:scount_model1}
 \end{center}
\end{figure}

While there is a steep slope 
across the 6--400~mJy 154~MHz Euclidean normalised differential source counts (see Fig.~\ref{fig:counts_MWA}), 
it is posited that this will flatten 
if there is a sizeable, fainter source population at $S_{154~\mathrm{MHz}} \lesssim 10$~mJy \citep{jackson2005}. This behaviour 
would mirror the flattening of the 1400~MHz source counts at $S_{1400} \lesssim 2$~mJy. Previous work 
has adopted a spectral index $\alpha \approx -0.7$ ($S \propto \nu^{\alpha}$), the canonical value for optically-thin 
synchrotron radiation, to extrapolate from 1400 to 154~MHz to predict the low frequency sky, but this could be na\"{i}ve: 
it assumes that the fainter population observed at 1400~MHz is indeed present at 154~MHz and has a typical spectral index of 
--0.7, and also that there is no low frequency source population with very steep spectra which is undetected at 1400~MHz. 

In fact, Appendix~A clearly demonstrates the problems inherent in using deep 1400~MHz catalogues
to predict the sky at a much lower frequency. The 154~MHz counts, which are well characterised
down to $\approx 6$~mJy, cannot be accurately predicted from 1400~MHz
counts using simple spectral index conversions. It is well known that a survey at higher/lower frequency preferentially detects
sources with flatter/steeper spectra. This selection bias, which was first analysed in detail by \cite{kellermann1964},
causes the effective spectral index distribution to change with frequency, which in turn renders simple extrapolations of counts 
invalid. 

For this reason, we chose not to extrapolate the 1400~MHz counts to 154~MHz to model the deep 154~MHz counts.
We explored three additional source count models, setting the source count slope below 6~mJy to 1.8 (model B), 2.0 (model C) and
2.2 (model D). Model A corresponds to the case explored above where there is no flattening in the counts below 6~mJy. The counts
are modelled as
\begin{eqnarray}
\label{eqn:scount_model_formula}
n(S) \equiv \frac{\mathrm{d}N}{\mathrm{d}S} \approx
\left\{
\begin{array}{ll}
k_{1} \left(\frac{S}{\mathrm{Jy}}\right)^{-\gamma_{1}}~\mathrm{Jy^{-1}~sr^{-1}}~\mathrm{for}~S_{\mathrm{low}} \leq S < S_{\mathrm{mid}} \\
k_{2} \left(\frac{S}{\mathrm{Jy}}\right)^{-\gamma_{2}}~\mathrm{Jy^{-1}~sr^{-1}}~\mathrm{for}~S_{\mathrm{mid}} \leq S \leq S_{\mathrm{high}}~\mathrm{.}
\end{array}
\right.
\end{eqnarray}
The values of the source count parameters $k_{1}$, $\gamma_{1}$, $k_{2}$, $\gamma_{2}$, $S_{\mathrm{low}}$, $S_{\mathrm{mid}}$ 
and $S_{\mathrm{high}}$ for each model are provided in Table~\ref{tab:scount_model_par}.

\begin{table*}
 \centering
 \caption{Parameter values adopted to model the 154~MHz counts with the resultant predicted classical and sidelobe confusion noise
assuming a beam size of 2.31~arcmin.}
 \label{tab:scount_model_par}
 \begin{tabular}{r r r r r r r r r r}
 \hline
 Model & $k_{1}$ & $\gamma_{1}$ & $k_{2}$ & $\gamma_{2}$ & $S_{\mathrm{low}}$ & $S_{\mathrm{mid}}$ & $S_{\mathrm{high}}$ & $\sigma_{\mathrm{c}}$ & $\sigma_{\mathrm{s}}$ \\
  &  &  &  &  & (mJy) & (mJy) & (mJy) & (mJy/beam) & (mJy/beam) \\
 \hline
 A & 6998 & 1.54 & 6998 & 1.54 & 0.1 & 6.0 & 400 & 1.7 & 3.5 \\
 B & 1841 & 1.800 & 6998 & 1.54 & 0.1 & 6.0 & 400 & 1.7 & 3.5 \\
 C & 661.8 & 2.000 & 6998 & 1.54 & 0.1 & 6.0 & 400 & 1.8 & 3.4 \\
 D & 237.9 & 2.200 & 6998 & 1.54 & 0.1 & 6.0 & 400 & 2.0 & 3.4 \\
 \hline
\end{tabular}
\end{table*}

Source count models A--D are shown in the top panel of Fig.~\ref{fig:scount_model1}.
The middle panel of Fig.~\ref{fig:scount_model1} shows the $P(D)$ distributions corresponding to these source count models.
The bottom panel of Fig.~\ref{fig:scount_model1} shows these distributions convolved with 
the system noise distribution and shifted to the left to remove offsets in the x-direction with respect to the observed $P(D)$ distribution.
Table~\ref{tab:scount_model_par} shows the predicted values of $\sigma_{\mathrm{c}}$
and $\sigma_{\mathrm{s}}$ for each source count model. $\sigma_{\mathrm{c}}$ and $\sigma_{\mathrm{s}}$ appear to be
relatively insensitive to the slope of the counts below 6~mJy. This indicates that sources below this flux density level
are too faint to contribute significantly to the confusion noise at this resolution. Table~\ref{tab:source_densities} shows the predicted source 
densities at 154~MHz above 5, 1, 0.5, 0.1 and 0.03~mJy/beam, for each source count model.

\begin{table*}
 \centering
 \caption{Predicted source densities at 154~MHz for various detection limits for each source count model explored in 
Section~\ref{Extending the observed source count}.}
 \label{tab:source_densities}
 \begin{tabular}{r r r r r}
 \hline
 $S_{\mathrm{lim}}$ / mJy/beam & N / $\mathrm{deg}^{2}$ & N / $\mathrm{deg}^{2}$ & N / $\mathrm{deg}^{2}$ & N / $\mathrm{deg}^{2}$ \\
  & for model A & for model B & for model C & for model D \\
 \hline
      5.00 &         62 &         62 &         63 &         63 \\
      1.00 &        157 &        190 &        224 &        268 \\
      0.50 &        231 &        320 &        425 &        580 \\
      0.10 &        560 &       1125 &       2038 &       3838 \\
      0.03 &       1077 &       2925 &       6742 &      16187 \\
 \hline
\end{tabular}
\end{table*}

\section{Discussion and future work}\label{Discussion}

We have analysed an MWA image of the EoR0 field at 154~MHz with 2.3~arcmin resolution to determine the noise contribution and behaviour 
of the source counts down to 30~mJy. The MWA EoR0 counts are in excellent 
agreement with the 7C counts by \cite{hales2007} and GMRT counts by \cite{williams2013}, \cite{intema2011} and \cite{ghosh2012};
our measurements are by far the most precise in the flux density range 30--200~mJy as a result of the large area of sky covered.
The differential GMRT counts are well represented by a power law of the form $\frac{dN}{dS} = 6998~S^{-1.54}  \, \mathrm{Jy}^{-1} \mathrm{sr}^{-1}$ 
between 6 and 400~mJy. Assuming no change in the slope of the 154~MHz counts
below 6~mJy, we estimate the classical confusion noise to be $\approx 1.7$~mJy/beam and the sidelobe
confusion noise to be $\approx 3.5$~mJy/beam; the predicted classical and
sidelobe confusion noise is relatively insensitive to the slope of the counts below 6~mJy.

Our $P(D)$ analysis suggests that, in this MWA EoR0 image, sidelobe confusion dominates other noise contributions despite the excellent $uv$ coverage.
This is a consequence of the large FoV and the huge number of detected sources.
We have identified three aspects of the data processing which potentially contribute to the sidelobe confusion
in these types of MWA images:
\begin{enumerate}

\item[(1)] The limited CLEANing depth. The snapshot images were CLEANed separately down to 100~mJy/beam before mosaicking, which is 22 
times the rms noise (4.5~mJy/beam) in the final mosaicked image. In practice, while CLEANing the image deeper is likely to 
lower the sidelobe confusion noise, ionospheric perturbations and primary beam-model errors introduce limitations in the ability
to deconvolve the image, making other approaches such as peeling more viable than deeper CLEANing.

\item[(2)] Far-field sources that have not been deconvolved: only the central 40 by 40~deg region of the image has been fully deconvolved, and
peeling limited to sources within 20~deg from the pointing centre. The importance of this effect will critically depend on how 
rapidly the rms of the MWA's synthesised beam decreases with distance from the beam centre.

\item[(3)] Source smearing due to the ionosphere. In each snapshot image, the ionosphere introduces a random displacement 
(typically 10--20~arcsec) in the source positions \citep{loi2015}. This smears out sources in 
the mosaicked image. Peeling corrects for the ionosphere whereas CLEANing does not. 

\end{enumerate}
It is unclear which of these factors is dominant; this will be the subject of further work.

Unlike previous estimates of the MWA classical confusion limit by \cite{thyagarajan2013b} and \cite{wayth2015},
our estimates do not rely on extrapolation of higher frequency counts, and we derive the exact shape
of the source $P(D)$ distribution. In Appendix~A, we show that the 154~MHz counts cannot be accurately reproduced from extrapolation of the
1400~MHz counts using simple spectral index conversions, demonstrating the need for deep source
counts at the same frequency as EoR observations rather than extrapolating from higher frequencies.

We plan to apply our $P(D)$ analysis to images from the GaLactic Extragalactic All-sky MWA 
\citep[GLEAM;][]{wayth2015} survey to assess how the different observing 
strategy and processing techniques affect sidelobe confusion. 
In so doing, we will quantify the magnitude of ionospheric smearing in detail.
The GLEAM survey covers the entire sky 
south of declination $+ 25^{\circ}$ at 72--231~MHz, reaching a sensitivity of $\approx 5$~mJy/beam.
We will also compare EoR specific imaging techniques to assess the impact of sidelobe confusion in detail, 
including Fast Holographic Deconvolution \citep{sullivan2012} and the Real-Time System \citep{mitchell2008,ord2009}.

Finally, we anticipate that the MWA will be upgraded to add a further 128 tiles, roughly doubling the current array resolution.
As a result, the classical confusion noise at 154~MHz will decrease by a factor of $\approx 5-10$ depending on the 
slope of the source counts below 6~mJy. The sidelobe levels are also expected to decrease, which will further
reduce sidelobe confusion. This raises the possibility of conducting large-area, sub-mJy continuum surveys, particularly
at the higher MWA observational frequency range ($\approx 200$~MHz).

\section*{Acknowledgments}\label{Acknowledgements}

This work makes use of the Murchison Radioastronomy Observatory, operated by CSIRO. We acknowledge the 
Wajarri Yamatji people as the traditional owners of the Observatory site. We thank the anonymous referee for their suggestions, which have improved this paper. CAJ thanks the Department of Science, Office of Premier \& Cabinet, WA 
for their support through the Western Australian Fellowship Program.  Support for the MWA comes from the U.S. 
National Science Foundation (grants AST-0457585, PHY-0835713, CAREER-0847753, and AST-0908884), the Australian 
Research Council (LIEF grants LE0775621 and LE0882938), the U.S. Air Force Office of Scientific Research 
(grant FA9550-0510247), and the Centre for All-sky Astrophysics (an Australian Research Council Centre of 
Excellence funded by grant CE110001020). Support is also provided by the Smithsonian Astrophysical Observatory, 
the MIT School of Science, the Raman Research Institute, the Australian National University, and the Victoria 
University of Wellington (via grant MED-E1799 from the New Zealand Ministry of Economic Development and an 
IBM Shared University Research Grant). The Australian Federal government provides additional support via the 
Commonwealth Scientific and Industrial Research Organisation (CSIRO), National Collaborative Research 
Infrastructure Strategy, Education Investment Fund, and the Australia India Strategic Research Fund, and 
Astronomy Australia Limited, under contract to Curtin University. We acknowledge the iVEC Petabyte Data Store, 
the Initiative in Innovative Computing and the CUDA Center for Excellence sponsored by NVIDIA at Harvard University, 
and the International Centre for Radio Astronomy Research (ICRAR), a Joint Venture of Curtin University and 
The University of Western Australia, funded by the Western Australian State government.

\setlength{\labelwidth}{0pt}


\appendix\section{Extrapolating the 1400~MH\lowercase{z} counts to predict the 154~MH\lowercase{z} sky}\label{Source count extrapolation}

\begin{figure*}
  \includegraphics[scale=0.57,angle=270]{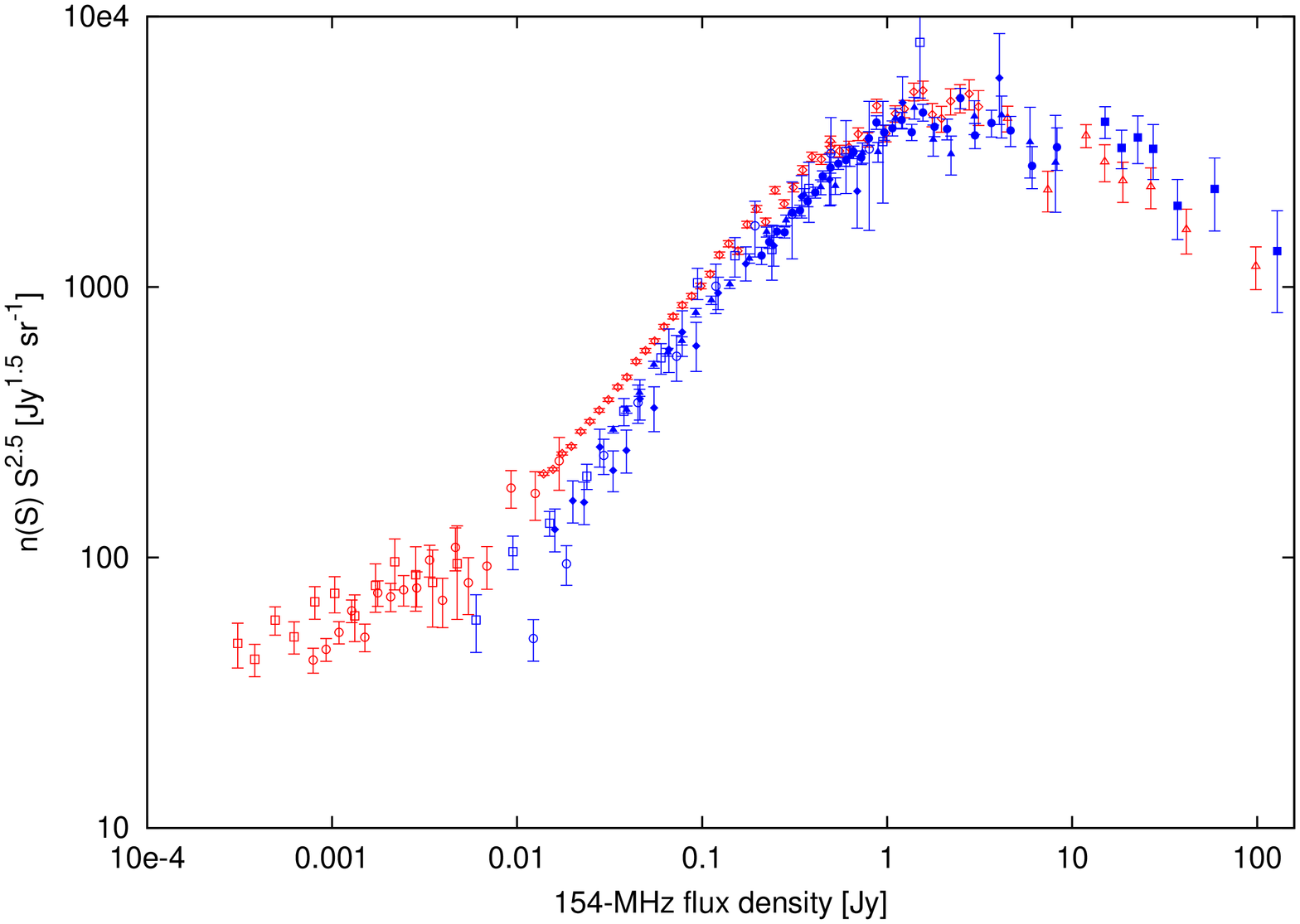}
  \caption{Comparison of counts in the frequency range 154--178~MHz extrapolated to 154~MHz (blue) with 1400~MHz counts
extrapolated to 154~MHz (red), in all cases assuming $\alpha = -0.75$. The data sources are as follows. 
154~MHz: filled circles, \citet{hales2007}; filled triangles, this paper; filled lozenges, \citet{williams2013}; 
open squares, \citet{intema2011}; open circles, \citet{ghosh2012};
178~MHz extrapolated to 154~MHz: filled squares, \citet{edge1959}; 1400~MHz extrapolated to 154~MHz: open circles, 
\citet{hales2014}; open squares, \citet{huynh2005}; open triangles, \citet{fomalont1974}; open lozenges, \citet{white1997}.}
  \label{fig:counts_154_1400_comparison1}
\end{figure*}

Fig.~\ref{fig:counts_154_1400_comparison1} shows counts in the frequency range 154--178~MHz extrapolated to 154~MHz
compared with 1400~MHz counts extrapolated to 154~MHz, in all cases assuming a spectral index of --0.75. 
It can be seen that extrapolation of the 1400~MHz counts to 154~MHz significantly overpredicts the 154~MHz counts below about 500~mJy. 
The density of sources at $S_{154} = 6$~mJy is overpredicted by about a factor of two.

\begin{figure*}
  \includegraphics[scale=0.57,angle=270]{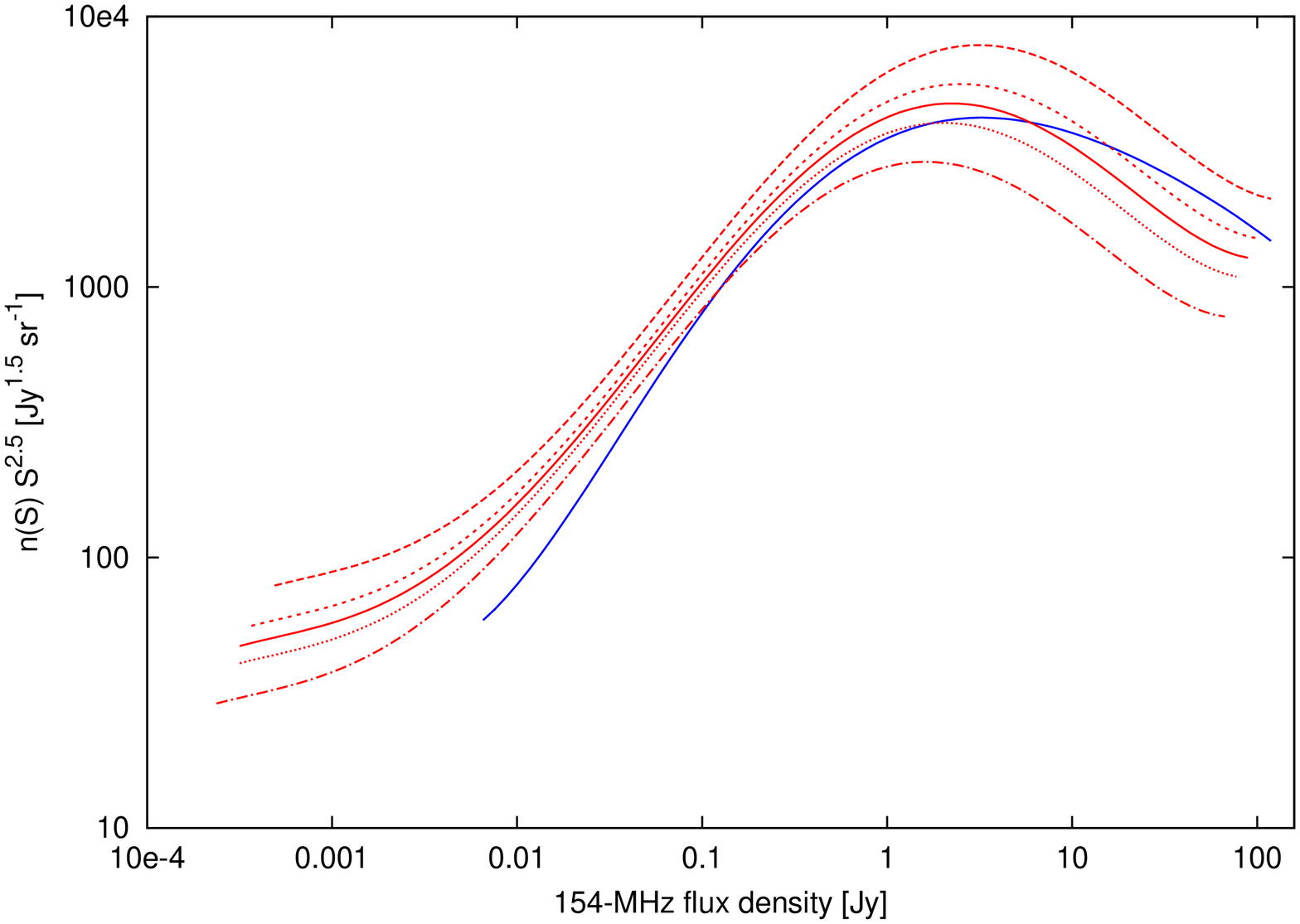}
  \caption{Polynomial fit to the 154~MHz counts (blue). Polynomial fit to the 1.4~GHz counts extrapolated
to 154~MHz assuming $\alpha = -0.90$ (red dashed line), $\alpha = -0.80$ (red hashed line), $\alpha = -0.75$ (red solid line),
 $\alpha = -0.70$ (red dotted line) and $\alpha = -0.60$ (red dot-dashed line).}
  \label{fig:counts_154_1400_comparison2}
\end{figure*}

Moreover, Fig.~\ref{fig:counts_154_1400_comparison2} shows that the 154~MHz counts above 6~mJy cannot 
be accurately reproduced from extrapolation of the 1.4~GHz counts using \textit{any} spectral index; the best fit
is obtained for a spectral index of --0.75. A polynomial fit to the 154~MHz counts is compared with a polynomial fit to the 1400~MHz 
counts extrapolated to 154~MHz assuming a spectral index of --0.90, --0.80, --0.75, --0.70 and --0.60. The integral of the 
squared difference between the two curves from $S_{154} = 6$~mJy to $S_{154} = 100$~Jy is minimised for $\alpha = -0.75$.

\label{lastpage}

\begin{thebibliography}{}


\bibitem[\protect\citeauthoryear{Beardsley et al.}{2013}]{beardsley2013} 
Beardsley~A.~P. et al., 2013, MNRAS, 429, L5 

\bibitem[\protect\citeauthoryear{Bernardi et al.}{2009}]{bernardi2009}
Bernardi~G. et al., 2009, A\&A, 500, 965

\bibitem[\protect\citeauthoryear{Bowman et al.}{2013}]{bowman2013} 
Bowman~J.~D. et al., 2013, PASA, 30, e031 


\bibitem[\protect\citeauthoryear{Bridle \& Schwab}{1999}]{bridle1999} 
Bridle~A.~H. \& Schwab~F.~R., 1999, Synthesis Imaging in Radio Astronomy II, 180, 371

\bibitem[\protect\citeauthoryear{Calabretta, Staveley-Smith \& Barnes}{2014}]{calabretta2014} 
Calabretta~M.~R., Staveley-Smith~L., Barnes~D.~G., 2014, PASA, 31, e007 

\bibitem[\protect\citeauthoryear{Condon}{1974}]{condon1974}
Condon~J., 1974, ApJ, 188, 279

\bibitem[\protect\citeauthoryear{Condon et al.}{1998}]{condon1998}
Condon~J.~J., Cotton~W.~D., Greisen~E.~W., Yin~Q.~F., Perley~R.~A.,
Taylor~G.~B., Broderick~J.~J., 1998, AJ, 115, 1693

\bibitem[\protect\citeauthoryear{Chapman et al.}{2012}]{chapman2012} 
Chapman~E. et al., 2012, MNRAS, 423, 2518 

\bibitem[\protect\citeauthoryear{Condon et al.}{2012}]{condon2012} 
Condon~J.~J. et al., 2012, ApJ, 758, 23 

\bibitem[\protect\citeauthoryear{Dewdney et al.}{2012}]{dewdney2012} 
Dewdney~P. et al., 2010, SKAMemo \#130, SKA Phase 1: Preliminary System Description, www.skatelescope.org/publications/

\bibitem[\protect\citeauthoryear{Edge et al.}{1959}]{edge1959} 
Edge~D.~O., Shakeshaft~J.~R., McAdam~W.~B., Baldwin~J.~E., Archer~S., 1959, MNRAS, 68, 37 

\bibitem[\protect\citeauthoryear{Fomalont, Bridle \& Davis}{1974}]{fomalont1974} 
Fomalont~E.~B., Bridle~A.~H., Davis~M.~M., 1974, A\&A, 36, 273 


\bibitem[\protect\citeauthoryear{Gehrels}{1986}]{gehrels1986} 
Gehrels~N., 1986, ApJ, 303, 336 

\bibitem[\protect\citeauthoryear{Ghosh et al.}{2012}]{ghosh2012} 
Ghosh~A., Prasad~J., Bharadwaj~S., Ali~S.~S., Chengalur~J.~N., 2012, MNRAS, 426, 3295 

\bibitem[\protect\citeauthoryear{Gower}{1966}]{gower1966} 
Gower~J.~F.~R., 1966, MNRAS, 133, 151

\bibitem[\protect\citeauthoryear{Hales et al.}{2007}]{hales2007}
Hales~S. et al., 2007, MNRAS, 382, 1639

\bibitem[\protect\citeauthoryear{Hales et al.}{2014}]{hales2014} 
Hales~C.~A. et al., 2014, MNRAS, 441, 2555

\bibitem[\protect\citeauthoryear{Hancock et al.}{2012}]{hancock2012}
Hancock~P. et al., 2012, MNRAS, 422, 1812

\bibitem[\protect\citeauthoryear{Harker et al.}{2010}]{harker2010} 
Harker~G. et al., 2010, MNRAS, 405, 2492 

\bibitem[\protect\citeauthoryear{Hopkins et al.}{2003}]{hopkins2003} 
Hopkins~A.~M., Afonso~J., Chan~B., Cram~L.~E., Georgakakis~A., Mobasher~B., 2003, AJ, 125, 465 

\bibitem[\protect\citeauthoryear{Hurley-Walker et al.}{2014}]{hurley-walker2014}
Hurley-Walker~N. et al., 2014, PASA, 31, e045

\bibitem[\protect\citeauthoryear{Huynh et al.}{2005}]{huynh2005} 
Huynh~M.~T., Jackson~C.~A., Norris~R.~P., Prandoni~I., 2005, AJ, 130, 1373 


\bibitem[\protect\citeauthoryear{Intema et al.}{2011}]{intema2011} 
Intema~H.~T., van Weeren~R.~J., R{\"o}ttgering~H.~J.~A., Lal~D.~V., 2011, A\&A, 535, A38 

\bibitem[\protect\citeauthoryear{Jackson \& Wall}{1999}]{jackson1999} 
Jackson~C.~A., Wall~J.~V., 1999, MNRAS, 304, 160 

\bibitem[\protect\citeauthoryear{Jackson}{2005}]{jackson2005}
Jackson~C., 2005, PASA, 22, 36


\bibitem[\protect\citeauthoryear{Kellermann}{1964}]{kellermann1964} 
Kellermann~K.~I., 1964, ApJ, 140, 969

\bibitem[\protect\citeauthoryear{Loi et al.}{2015}]{loi2015} 
Loi~S.~T. et al., 2015, Radio Science, 50, 574 

\bibitem[\protect\citeauthoryear{Longair}{1966}]{longair1966} 
Longair~M.~S., 1966, MNRAS, 133, 421

\bibitem[\protect\citeauthoryear{Lonsdale et al.}{2009}]{lonsdale2009} 
Lonsdale~C.~J. et al., 2009, Proc. IEEE, 97, 1497 

\bibitem[\protect\citeauthoryear{Mauch et al.}{2003}]{mauch2003} 
Mauch~T., Murphy~T., Buttery~H.~J., Curran~J., Hunstead~R.~W., Piestrzynski~B., Robertson~J.~G., Sadler~E.~M., 
2003, MNRAS, 342, 1117 


\bibitem[\protect\citeauthoryear{Mitchell \& Condon}{1985}]{mitchell1985} 
Mitchell~K.~J., Condon~J.~J., 1985, AJ, 90, 1957 

\bibitem[\protect\citeauthoryear{Mitchell et al.}{2008}]{mitchell2008} 
Mitchell~D.~A., Greenhill~L.~J., Wayth~R.~B., Sault~R.~J., Lonsdale~C.~J., Cappallo~R.~J., Morales~M.~F., Ord~S.~M., 2008, IEEE Journal of Selected Topics in Signal Processing, 2, 707 

\bibitem[\protect\citeauthoryear{Morales \& Hewitt}{2004}]{morales2004} 
Morales~M.~F., Hewitt~J., 2004, ApJ, 615, 7 

\bibitem[\protect\citeauthoryear{Offringa et al.}{2014}]{offringa2014} 
Offringa~A.~R. et al., 2014, MNRAS, 444, 606 

\bibitem[\protect\citeauthoryear{Offringa et al.}{2015}]{offringa2015} 
Offringa~A.~R. et al., 2015, PASA, 32, 8 

\bibitem[\protect\citeauthoryear{Offringa et al.}{2016}]{offringa2016} 
Offringa~A.~R. et al., 2016, MNRAS, submitted

\bibitem[\protect\citeauthoryear{Ord et al.}{2009}]{ord2009} 
Ord~S., Greenhill~L., Wayth~R., Mitchell~D., Dale~K., Pfister~H., Edgar~R., 2009, ASPC, 411, 127 

\bibitem[\protect\citeauthoryear{Parsons et al.}{2014}]{parsons2014} 
Parsons~A.~R. et al., 2014, ApJ, 788, 106 

\bibitem[\protect\citeauthoryear{Scheuer}{1957}]{scheuer1957} 
Scheuer~P.~A.~G., 1957, Proceedings of the Cambridge Philosophical Society, 53, 764 

\bibitem[\protect\citeauthoryear{Sullivan et al.}{2012}]{sullivan2012} 
Sullivan~I.~S. et al., 2012, ApJ, 759, 17 

\bibitem[\protect\citeauthoryear{Thyagarajan et al.}{2013a}]{thyagarajan2013a} 
Thyagarajan~N. et al., 2013a, ApJ, 776, 6 

\bibitem[\protect\citeauthoryear{Thyagarajan}{2013b}]{thyagarajan2013b}
Thyagarajan~N., 2013b, Notes on MWA 128T confusion limits, MWA Technical Memo. Available at: http://mwa-lfd.haystack.mit.edu/knowledgetree/view.php?fDocumentId=826

\bibitem[\protect\citeauthoryear{Tingay et al.}{2013}]{tingay2013} 
Tingay~S.~J. et al., 2013, PASA, 30, e007 

\bibitem[\protect\citeauthoryear{Toffolatti et al.}{1998}]{toffolatti1998} 
Toffolatti~L., Argueso Gomez~F., de Zotti~G., Mazzei~P., Franceschini~A., Danese~L., Burigana~C., 1998, MNRAS, 297, 117 

\bibitem[\protect\citeauthoryear{Trott, Wayth \& Tingay}{2012}]{trott2012} 
Trott~C.~M., Wayth~R.~B., Tingay~S.~J., 2012, ApJ, 757, 101 

\bibitem[\protect\citeauthoryear{van Haarlem et al.}{2013}]{van_haarlem2013} 
van~Haarlem~M.~P. et al., 2013, A\&A, 556, A2 


\bibitem[\protect\citeauthoryear{Wall}{1994}]{wall1994} 
Wall~J.~V., 1994, AuJPh, 47, 625 

\bibitem[\protect\citeauthoryear{Wayth et al.}{2015}]{wayth2015} 
Wayth~R.~B. et al., 2015, PASA, 32, e025 

\bibitem[\protect\citeauthoryear{White et al.}{1997}]{white1997} 
White~R.~L., Becker~R.~H., Helfand~D.~J., Gregg~M.~D., 1997, ApJ, 475, 479 

\bibitem[\protect\citeauthoryear{Wieringa}{1991}]{wieringa1991} 
Wieringa~M., 1991, PhD thesis, Leiden University

\bibitem[\protect\citeauthoryear{Williams, Intema \& R\"{o}ttgering}{2013}]{williams2013}
Williams~W.~L., Intema~H.~T., R\"{o}ttgering~H.~J.~A., 2013, A\&A, 549, A55


\bibitem[\protect\citeauthoryear{Zwart et al.}{2015}]{zwart2015} 
Zwart~J. et al., 2015, in Bourke~T.~L. et al., eds, Proc. Advancing Astrophysics with the Square Kilometre Array, Astronomy below the Survey Threshold, id.~172. Available at: http://pos.sissa.it/cgi-bin/reader/conf.cgi?confid=215\#session-2110

\end{thebibliography}
\end{document}